\documentclass[12pt]{article}
\usepackage[english]{babel}
\usepackage{amsfonts}
\usepackage{amssymb}
\usepackage{latexsym}
\usepackage{graphicx}
\usepackage{epsfig}
\usepackage{epsf}

\newcommand{\sectiono}[1]{\section{#1}\setcounter{equation}{0}}

\def\Gtilde{{\widetilde\Gamma}}
\def\Qtilde{{\widetilde Q}}
\def\Ltilde{{\widetilde L}_0}
\def\Lhat{{\widehat L}_0}
\def\Dtilde{{\widetilde D}}
\def\Dcaltilde{\widetilde{\cal D}}
\def\Dhat{{\widehat D}}
\def\Mhat{{\widehat M}}
\def\Zhat{{\widehat Z}}
\def\be{\begin{equation}}
\def\ee{\end{equation}}
\def\bea{\begin{eqnarray}}
\def\eea{\end{eqnarray}}
\def\sss{\scriptscriptstyle}

\def\brs{\delta_{\rm BRS}}
\def\ss#1{{\sss (#1)}}
\def\Vtilde#1{{\widetilde V}_{\sss #1}}
\def\F#1{{F}_{\sss #1}}

\def\Wtilde#1{{\widetilde W}_{\sss #1}}
\def\Wcheck#1{\check{W}_{\sss #1}}

\def\htilde#1{{\tilde{\mathrm{h}}}^\ss{#1}}
\def\H#1{{\cal H}^\ss{#1}}
\def\hcheck#1{{\check{\mathrm{h}}}^\ss{#1}}

\def\n#1{{\tilde\mathrm n}^{\sss (#1)}}
\def\ncheck#1{{\check\mathrm{n}}^{\sss (#1)}}
\def\O#1{\Omega_{\sss #1}}

\def\Zcheck{\check{Z}}

\begin{document}

\begin{titlepage}
\rightline{GEF-TH-17-2006 }
\vskip 1in
\begin{center}
\def\thefootnote{\fnsymbol{footnote}}
{\large The Spectrum of Open String
Field Theory\\  at the Stable Tachyonic Vacuum}
\vskip 0.3in
C. Imbimbo\footnote{E-Mail: camillo.imbimbo@ge.infn.it} 
\vskip .2in
{\em Dipartimento di Fisica, Universit\`a di Genova\\
and\\ Istituto Nazionale di Fisica Nucleare, Sezione di Genova\\
via Dodecaneso 33, I-16146, Genoa, Italy}
\end{center}
\vskip .4in
\begin{abstract}
We present a level (10,30) numerical computation of the spectrum of
quadratic fluctuations of Open String Field Theory around the tachyonic vacuum, both in the scalar and in the vector sector. Our results are consistent with Sen's conjecture about gauge-triviality 
of the small excitations. The computation is sufficiently 
accurate to provide robust evidence for the absence of the photon from the open string spectrum. We also observe that ghost string 
field propagators develop double poles. We show that this requires non-empty BRST cohomologies at non-standard ghost numbers.  
We comment about the relations of our results with recent work on the same subject.

\end{abstract}
\vfill
\end{titlepage}
\setcounter{footnote}{0}

\sectiono{Introduction, Summary and Discussion}

In this paper we extend and improve the analysis of bosonic Open String
Field Theory (OSFT) at the stable vacuum that we started in a previous
work \cite{Giusto:2003wc}.  OSFT possesses a classical, translational
invariant solution whose energy density exactly cancels the brane tension
and which is thought to represent the closed string vacuum with no open
strings \cite{sen1,sen2}. The existence of such a solution has been
persuasively demonstrated first \cite{sz,mt,gr} within the level
truncation (LT) expansion \cite{ks}, and, more recently, analytically
\cite{Schnabl:2005gv}.  The closed string interpretation requires that the
spectrum of quadratic fluctuations around this classical solution be not
only tachyon-free but also gauge-trivial. This expected property of the
tachyonic vacuum goes under the name of Sen's OSFT third conjecture.

In \cite{Giusto:2003wc} we explained what Sen's third conjecture implies
for the gauge-fixed quadratic OSFT action expanded around the stable
vacuum: {\it each pole} of the open string field propagator should cancel
with appropriate poles of the (second quantized) ghost string fields.  To
state it more precisely, let us denote by $\Ltilde^\ss{n}(p)$ (where
$n=0,1,\ldots$ is (minus) the second quantized ghost number\footnote{We
are adopting the convention in which the $SL(2,\mathbb{R})$ invariant
vacuum $|0\rangle$ has ghost number -1.  In the natural hermitian product,
$\Ltilde^\ss{-n}=(\Ltilde^\ss{n})^\dagger$.  Thus ghost and antighost
string fields form canonically conjugate pairs $(\phi_{-n}, \phi_{n})$
with $n>0$.  })  the gauge-fixed kinetic operators for both matter and
ghost string fields acting on states of momentum $p$ and ghost number
$n$. $\Ltilde^\ss{n}(p)$ is the restriction to states of ghost number $n$
and momentum $p$ of the operator
\be
\Ltilde =\{ \Qtilde, b_0\}
\label{kineticoperator}
\ee
where $\Qtilde$ is the BRS operator associated with the classical stable
OSFT solution and $b_0$ is the zero mode of the 2d CFT antighost field
that implements the Siegel gauge condition.  If $\det \Ltilde^\ss{n}(p)$
has a zero of order $d_n$ for $p^2=-m^2$, the number of physical ---
i.e. {\it gauge-invariant} --- degrees of freedom of mass $m$ is given by
the Fadeev-Popov index:
\be
I_{FP}(m) = d_0 - 2\, d_1 + 2\, d_2 +\cdots = \sum_{n=-\infty}^\infty 
(-1)^n\, d_n
\label{fpindex0}
\ee
The spectrum of gauge-invariant quadratic fluctuations is empty if, and
only if, the above index vanishes for all $p^2$. 

Both \cite{Giusto:2003wc} and the present paper study the spectrum of
quadratic fluctuations of OSFT around the stable vacuum within the
framework of the LT expansion. The key drawback of LT expansion is that it
breaks (second quantized) BRS invariance of the gauge-fixed OSFT action
around the stable vacuum.  Consequently, poles of propagators of matter
and ghost string fields which are degenerate in the exact theory
correspond, in the level truncated theory, to multiplets of poles that are
only approximately degenerate.  The Fadeev-Popov index (\ref{fpindex0})
should therefore be defined, in the level truncated theory, by including
zeros of gauge-fixed kinetic operator $\Ltilde^\ss{n}(p)$ which belong to
the same approximately degenerate multiplet. In order for this definition
to make sense the level has to be large enough that the splitting among
zeros belonging to the same multiplet is significantly smaller than the
separation between multiplets. It is expected --- and explicit numerical
computations confirm this --- that matter and ghost propagators poles
begin clustering together into well-defined approximately degenerate
multiplets for levels which are increasingly large as $m^2= -p^2 \to
\infty$.  In practice, therefore, the level truncated numerical analysis
can probe reliably only a limited range of values of $m^2$.

The analysis of \cite{Giusto:2003wc} was limited to the Lorentz {\it
scalar} sector of the theory. Numerical computations were performed using
the approximation which, in terminology of \cite{sz}, was of type $(L,
3L)$ for levels $L$ up to 6 and of type $(L,2L)$ for levels $L=7,8,9$.
For $p^2 > -6.0$ propagators poles were found only in the twist-odd
sector: it was observed that they form an approximately degenerate
multiplet with vanishing Fadeev-Popov index around
$-p^2=m^2_{scalar,-}\approx 2.1$. In the twist-even scalar sector,
propagators have no poles up to $-p^2 \approx 6$: at the level reached by
the computation, poles with $p^2 < -6.0$ do not show yet any clear and
stable multiplet structure.

Although these findings are well consistent with gauge-triviality of the
spectrum of quadratic excitations, an unexpected result was also obtained
in \cite{Giusto:2003wc}. The non-vanishing Fadeev-Popov degrees $d_n$ of
the approximately degenerate multiplet of zeros of $\det\Ltilde^\ss{n}(p)$
at $p^2=-m^2_{scalar,-}$ were found to be
\be
d_0=2\qquad d_1=2\qquad d_2=1
\label{fpdegrees}
\ee
It was moreover observed that while the (approximate) double zero of the
determinant of the matter kinetic operator is associated to {\it two}
distinct vanishing eigenvalues of $\Ltilde^\ss{0}(p)$, the double zero of
$\det\Ltilde^\ss{\pm1}(p)$ is due to a {\it single} eigenvector of the
kinetic operator of ghost numbers $\pm1$ whose eigenvalue
\be
\lambda(p^2)\propto (p^2+m^2_{scalar,-})^2
\label{doublepole}
\ee
has an (approximate) {\it double} zero at $p^2=-m^2_{scalar,-}$.  In other
words there exists a ghost-antighost string field pair with ghost number
$\pm1$ whose propagator develops a double pole for $p^2=-m^2_{scalar,-}$.

Physical states are elements of the $\Qtilde$-cohomology with zero ghost
number.  Let us denote by $\H{n}(\Qtilde)$ the cohomology of $\Qtilde$ on
states of ghost number $n$. Because of (\ref{kineticoperator}), the zeros
of the gauge-fixed kinetic operators also encode properties of $\Qtilde$
acting on $\H{n}(\Qtilde)$ with $n$ different than zero. In
\cite{Giusto:2003wc} it was argued that the double pole of the propagator
of the ghost-antighost string field pair at $p^2=-m^2_{scalar, -}$,
although consistent with the vanishing of $\H0(\Qtilde)$, requires as well
that 
\be 
\dim \H{-1}(\Qtilde)= \dim \H{-2}(\Qtilde)=1
\label{dimcoho}
\ee 
for the same value of $p^2$.
 
\begin{figure}[p]
\begin{center}
\includegraphics[scale=.8, clip=false]{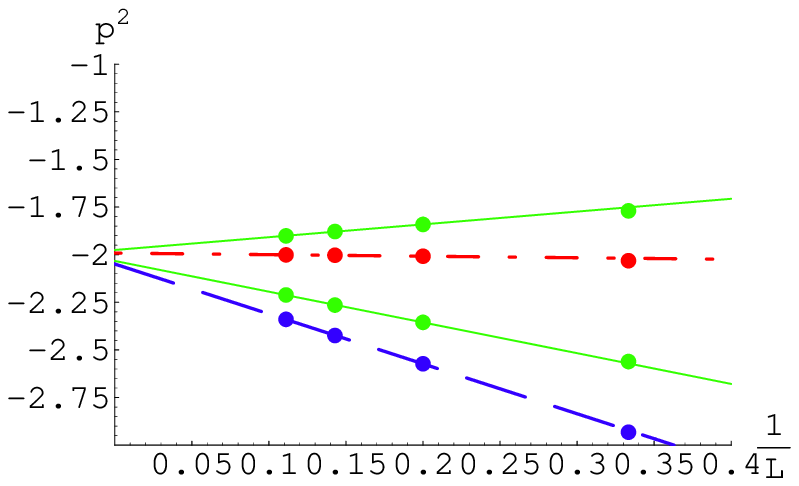}
\end{center}
\caption[x] {\footnotesize The first group of zeros of
$\Delta^\ss{n}_{-}(p^2)$ in the scalar odd sector at $p^2\approx -2.0$ for levels $L=3, 5, 7, 9$. $J=0$ dot-dashed-red, $J=1/2$ solid-green, $J=1$ dashed-blue.}
\label{f:scalarodd}
\begin{center}
\includegraphics[scale=.8, clip=false]{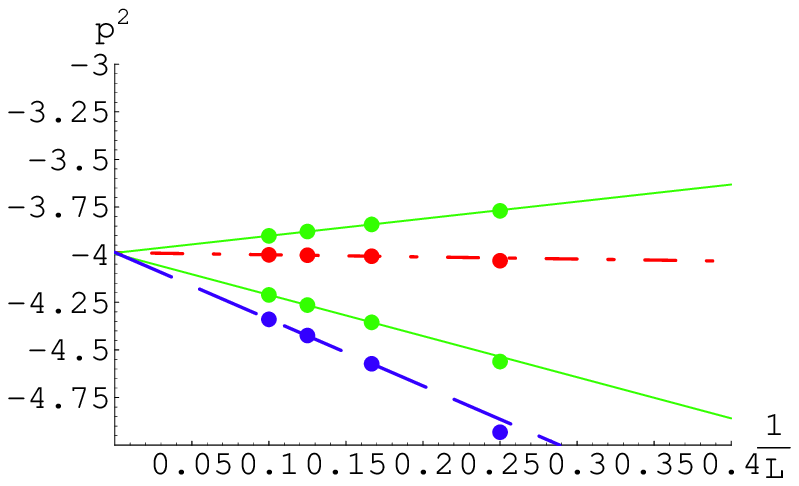}
\end{center}
\caption[x] {\footnotesize The first group of zeros of
$\Delta^\ss{n}_{+}(p^2)$ in the vector even sector at $p^2\approx -4.0$ for levels $L= 4, 6, 8, 10$. $J=0$ dot-dashed-red, $J=1/2$ solid-green, $J=1$ dashed-blue.}
\label{f:vectoreven}
\begin{center}
\includegraphics[scale=0.7, clip=false]{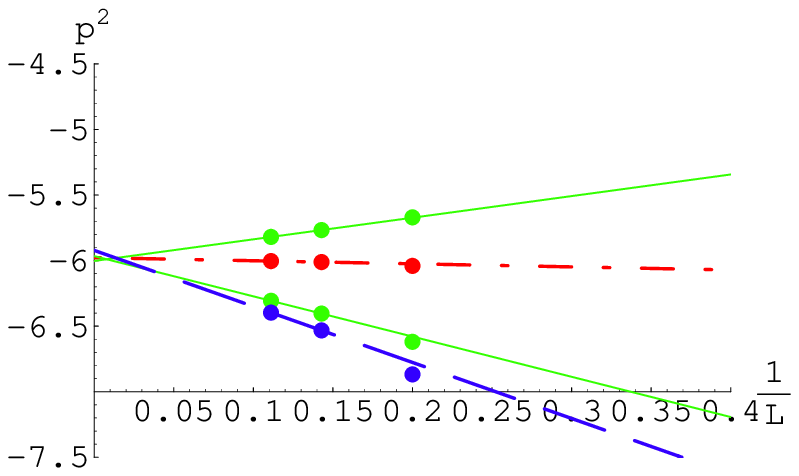}
\end{center}
\caption[x] {\footnotesize The first group of zeros  of 
$\Delta^\ss{n}_{-}(p^2)$ in the vector odd sector
at $p^2\approx -6.0$ for levels $L= 5, 7, 9$. $J=0$ dot-dashed-red, $J=1/2$ solid-green, $J=1$ dashed-blue.}
\label{f:vectorodd}
\end{figure}

In the present paper we confirm and refine this analysis in various
ways. First, we compute the gauge-fixed kinetic operators
$\Ltilde^\ss{n}(p)$ for both the {\it scalar} and the {\it vector} Lorentz
sector.  We also improve our LT computation, by performing the numerical
evaluations in the $(L, 3L)$ approximation\footnote{
Our numerical findings confirm the conclusion, 
shared by other authors, that the approximation of type (L, 2L) 
is not satisfactory for this problem.} up to level 
$L=10$. We find, on top of the multiplet of poles of the twist-odd scalar
propagators already discovered in \cite{Giusto:2003wc}, multiplets of
propagator poles, which are approximately degenerate and have vanishing
Fadeev-Popov indices, both in the twist-even and in the twist-odd {\it
vector} sector.  The increased level reached by the computation allows for
a simple linear extrapolation of the locations of the poles of the
approximately degenerate Fadeev-Popov multiplets. The poles of propagators
of different ghost numbers when extrapolated at level $L=\infty$ become
indeed nearly coincident: See Table~\ref{tabellazeri} and 
Figures~\ref{f:scalarodd}-\ref{f:vectorodd} of
 Section~\ref{sec:numerical}. The extrapolated values 
of the masses of the degenerate multiplets are
\bea
&&m^2_{scalar,-} = -p^2 =2.0\nonumber\\
&& m^2_{vector, +} = -p^2 =4.0\nonumber\\
&&m^2_{vector, -} = -p^2 =5.9
\label{poles}
\eea
where the indices $\pm$ refers to the twist even/odd sectors.  The values
(\ref{poles}) of the poles are intriguingly consistent with {\it integer}
even values of $m^2$.  The observed structure of the three multiplets is
the same. Fadeev-Popov degrees are as in Eq. (\ref{fpdegrees}); in all the
three cases listed in (\ref{poles}), there exists a {\it single}
ghost-antighost string field pair with ghost numbers $\pm1$ whose
propagator develops an (approximate) {\it double} pole.

These numerical results confirm Sen's conjecture for both scalars and
vectors in the region $p^2 \gtrsim -6$. In particular our approximation
should be quite accurate for $p^2=0$: hence, our computation provides (the
first) robust direct evidence for the absence of the photon in the
non-perturbative stable vacuum.

At the same time, the arguments developed in \cite{Giusto:2003wc} imply
that non-empty cohomologies at non-standard ghost numbers must exist in
the vector sectors as well as in the scalar sector.  Since this conclusion
is somewhat unexpected and appears to contradict other works
\cite{rsz},\cite{efhm},\cite{Ellwood:2006ba}, we reconsider and strengthen
in the present paper the analysis of \cite{Giusto:2003wc}, which
relates the double pole of the ghost-antighost string field pair to
non-vanishing cohomologies at non-standard ghost numbers.

The main mathematical tool used in this analysis is the long
infinite cohomology sequence
\be
\cdots\; {\mathop{\longrightarrow}}\; \htilde{n}(\Qtilde)\;
{\mathop{\longrightarrow}}\;\H{n}(\Qtilde)\; {\mathop{\longrightarrow}}
\;\hcheck{n-1}(\Qtilde)\;{\mathop{\longrightarrow}}\;\htilde{n+1}
(\Qtilde)\;{\mathop{\longrightarrow}}\;\cdots
\label{nonpertbottsequenceintro}
\ee
that computes the cohomologies $\H{n}(\Qtilde)$ in terms of {\it relative}
tilde $\htilde{n}(\Qtilde)$ and check $\hcheck{n}(\Qtilde)$
$\Qtilde$-cohomologies.  Both tilde and check relative cohomologies are
defined on fields which satisfy the Siegel gauge condition.  The tilde
relative $\htilde{n}$ cohomology is the cohomology of $\Qtilde$ on fields
$\phi_n$ that are in the kernel of both $\Ltilde$ and $b_0$
\be
b_0\, \phi_n = \Ltilde\, \phi_n =0
\label{relativecondition}
\ee 
This is a consistent cohomological problem since $ \Ltilde = \{\Qtilde
, b_0\}$.  In \cite{Giusto:2003wc} we also introduced the check relative
$\Qtilde$-cohomology $\hcheck{n}$ on the space of fields that satisfy both
(\ref{relativecondition}) and
 \be
 {\hat Z} \, \phi_n = \Lhat\,\phi_{n+1}
 \ee
where the operators  $\Zhat$ and $\Lhat$ are defined by the 
decomposition of
 $\Qtilde$ in the $c_0$ and $b_0$ algebra
\be
\Qtilde = c_0\, \Lhat + b_0\,\Dhat + \Mhat  + c_0\, b_0\, \Zhat
\label{qtildedecomposition}
\ee
We will review in Section \ref{sec:relabscohomologies} why this also is a
well posed cohomological problem for $\Qtilde$.  Since $\Zhat = [c_0,
\Ltilde]$, the BRS operator associated with the unstable ``perturbative''
vacuum represented by the bosonic 25-brane has $\Zhat=0$. Consequently,
the tilde and the check cohomologies of the ``perturbative'' BRS operator
coincide.  Our numerical computations indicate that $\Zhat$ does not
vanish for the non-perturbative $\Qtilde$.

In \cite{Giusto:2003wc} the cohomology sequence
(\ref{nonpertbottsequenceintro}) was derived by making certain technical
assumptions that were not otherwise proven. It was assumed that the space
of gauge-fixed states decomposes as the direct sum of the kernel and the
image of $\Ltilde$. In this paper we relax this hypothesis and establish
the validity of (\ref{nonpertbottsequenceintro}) beyond reasonable
doubt. We emphasize that this is an {\it exact} result, independent of any
numerical computation.

We also show, without invoking any of the mathematical structures provided
by the technical assumption made in \cite{Giusto:2003wc}, that the
``experimentally'' observed propagator double poles of the ghost
string field pairs with ghost number $\pm1$ are compatible with the exact
long sequence only if Eq.  (\ref{dimcoho}) for non-standard cohomologies
holds for the corresponding values of $p^2$ listed in (\ref{poles}). Furthermore we refine this
result by making use of the following observation.\footnote{In
\cite{Giusto:2003wc} it was established that a certain number of different
assignments for the relative cohomologies, all of which implied
(\ref{dimcoho}), were consistent with the long exact sequence. The
analysis could not provide a unique possibility for the explicit
representatives of the absolute cohomologies. This is achieved in the present paper.} It is known
\cite{Zwiebach:2000vc} that $\Ltilde$ commutes with an $SU(1,1)$ symmetry
whose $J_3$ generator is half of the ghost number. Although the full
$SU(1,1)$ symmetry does not commute with $\Qtilde$, the $J_+$ generator
does.  Exploiting this symmetry we show that the null vectors of the
kinetic operators for ghost number -1 and -2 provide explicit
representatives of the non-vanishing cohomologies:
\bea
&&\Ltilde^\ss{-2}(p)\,v_{-2} =0\qquad [v_{-2}] \in \H{-2}(\Qtilde)
\nonumber\\
&&\Ltilde^\ss{-1}(p)\,v_{-1}=0\qquad [v_{-1}] \in \H{-1}(\Qtilde)
\label{exoticcoho}
\eea

We provide the detailed proof of (\ref{exoticcoho}) in Section
\ref{sec:qaction}.  However it is useful to summarize here the basic
reasons for this result.  The important observation is that $\Qtilde$ acts
on the space ${\widetilde W}(p)$ spanned by the vectors which are in the
kernel of both $b_0$ and $\Ltilde$ at a given $p^2$. These are the null
vectors responsible for the poles of the string field propagators.  The
approximately degenerate poles that we found numerically correspond in the
exact theory to a null space ${\widetilde W}(p)$ with the same dimension
six for all the three sectors and values of $p^2$ listed in
Eq. (\ref{poles}).  For these values of $p^2$, $\Qtilde$ acting on
${\widetilde W}(p)$ reduces to the operator $\Mhat$ that appears in the
decomposition (\ref{qtildedecomposition}).  Therefore $\Mhat^2=0$ on the
six-dimensional space ${\widetilde W}(p)$. The (Witten) index of this
supersymmetry is $4-2=2$ and equals the index of the relative tilde and
check cohomologies
\be
\mathrm{index}\, \tilde{\mathrm{h}} =\sum_{n} (-1)^n \dim \htilde{n} =
\mathrm{index}\, \check{\mathrm{h}} =\sum_{n} (-1)^n \dim \hcheck{n}=2
\ee
This implies that the relative cohomologies cannot all vanish. 

${\widetilde W}(p)$ decomposes into a singlet, a doublet and a triplet
representation of the $SU(1,1)$ symmetry.  Clearly, there exists only a
finite number of inequivalent representations of the supersymmetry
operator $\Mhat$ acting on the finite-dimensional space ${\widetilde
  W}(p)$.  Among these representations one should focus on those for which
$J_+$ commutes with $\Mhat$. There are four such inequivalent
representations, as shown in Section \ref{sec:qaction}.  Since
${\widetilde W}(p)$ is finite dimensional, the infinite long exact
sequence becomes a finite exact sequence.  The fact that
$\H{0}(\Qtilde)=0$ causes the exact sequence to split into four shorter
sequences, putting further constrains on the possible representations of
$\Mhat$. In the end, it becomes a matter of simple linear algebra to show
that only one representation of $\Mhat$ is compatible with the exact
sequences: 
\be 
\Mhat\, v_{\pm 2} =0\qquad\Mhat\, v_{\pm1} =0\qquad \Mhat\,
v_0^{t}=0\qquad \Mhat\, v_0^{s} = v_1
\label{mhataction}
\ee 
where $v_0^s$ is the $SU(1,1)$ singlet, $\{v_{\pm 1}\}$ the doublet,
and $\{v_0^t, v_{\pm 2}\}$ the triplet.

We went into these details to clarify that the result (\ref{exoticcoho})
relies exclusively on the assumption that the multiplets of approximately
degenerate propagators poles that we found numerically for the values of
$p^2$ listed in (\ref{poles}) do really correspond in the {\it exact}
theory to multiplets of exactly degenerate poles.  This assumption cannot
of course rigorously be proven by numerical methods alone.  {\it A priori}
one can imagine that, as the level increases, either some of the poles we
found disappear or some new pole shows up.  This, although possible in
principle, seems however unlikely for the following reasons.

To start with, the zeros $p^2_{n}(L)$ of the determinants
$\det\Ltilde^{\ss n}(p)$ as the level is increased from $L=4$ to $L=10$
are nicely interpolated by linear relations
\be
p^2_{n}(L) = p^2_n +  {q_n\over L} 
\ee
with intercepts $p^2_n$ (corresponding to $L=\infty$) which are
independent of the ghost numbers $n=0,1,2$ with remarkably good 
approximation: See Table~\ref{tabellazeri} and Figures~\ref{f:scalarodd}-
\ref{f:vectorodd} of Section \ref{sec:numerical}.  The zeros
with $n=0,2$ correspond to eigenvalues which vanish linearly in $p^2$: on
topological grounds, they are stable and therefore unlikely to disappear
altogether.  As we have mentioned above, the zeros with $n=1$ correspond
instead to eigenvalues which vanish quadratically in $p^2$ and thus they
are not protected by topological reasons. Therefore one could think that
--- contrary to what our numerical computations seem to indicate --- the
pair of zeros with $n=1$ do not correspond, in the exact theory, to a
single eigenvalue with a double zero: rather, one might fear that, as the
level is increased, this pair of zeros would eventually be lifted.
However, if this were the case, the Fadeev-Popov index would become
positive and physical degrees of freedom would appear. In other words, if
we {\it assume} Sen's conjecture, it becomes very difficult to provide an
interpretation of our numerical findings different than what we have
proposed.

Two more independent tests support the conclusion about the exotic
cohomologies of $\Qtilde$ that we inferred from the numerical
computation. The first argument was already developed in
\cite{Giusto:2003wc}.  If $\Qtilde(p)$ has non-vanishing cohomologies for
discrete values of $p^2$ only and $\H{0}(\Qtilde)=0$, then the following
equation must hold in the {\it exact} theory
\be
0= \dim \H{-1}(\Qtilde) -  \dim \H{-2}(\Qtilde)+\cdots
\ee  
Our result (\ref{exoticcoho}) does satisfy this constraint and this seems
to be a non-trivial check of its correctness.

One more reasoning that strengthen our belief in the correctness of our
conclusion (\ref{exoticcoho}) is suggested by the numerical extrapolations
(\ref{poles}) obtained in the present paper.  Poles of propagators that
correspond to gauge-trivial excitations are, in general, gauge-dependent.
According to our extrapolations, the poles of the string fields seem to
correspond, in the exact theory, to {\it integer} even values of $m^2$.
It becomes difficult to understand why this is so unless such poles have a
{\it gauge-invariant} meaning, like the one provided by (\ref{dimcoho}).

After we completed our numerical computations, the paper
\cite{Ellwood:2006ba} appeared where, following previous work \cite{efhm},
an analytical proof of the absence of physical states of OSFT around the
stable vacuum is presented. The idea of this proof is to show that the
identity state $\mathcal{I}$ is $\Qtilde$-exact:
\be
\mathcal{I}= \Qtilde\, A
\label{istate}
\ee
The authors of \cite{Ellwood:2006ba} provide an explicit expression for
the trivializing state $A$
\be
A = {1\over \mathcal{L}_0} \mathcal{B}_0\mathcal{I}
\label{astate}
\ee
where the operators $ \mathcal{L}_0$ and $ \mathcal{B}_0$ are obtained
from the usual perturbative operators $L_0$ and $b_0$ by means of a
certain coordinate transformation. The exactness of the identity implies
not only the emptiness of $\H{0}(\Qtilde)$ but also that of
$\H{n}(\Qtilde)$ for $n$ generic: this of course contradicts our result in
(\ref{exoticcoho})\footnote{The conflict between our result about
cohomologies at non-zero ghost number and the triviality of the identity
prompted us to both improve our numerical results and to relax the
technical hypothesis that were assumed in \cite{Giusto:2003wc} to
derive the exact long sequence (\ref{nonpertbottsequenceintro})}.

We do not have yet a definite understanding of this conflict. The
discussion we presented above makes it clear that the only reasonable way
to avoid our conclusion regarding cohomologies at non-standard ghost
numbers is that level truncation is simply not appropriate to study the
spectrum of OSFT at $p^2<0$, regardless of the level.  In other words one
has to admit that the approximately degenerate multiplets of poles that we
detected are a finite level artifact that have no correspondence in the
exact theory. We just reviewed the reasons why this, although possible in
principle, seems difficult to understand.

On the other hand, it should be remarked that the proof presented in
\cite{Ellwood:2006ba} has a somewhat formal character. The reason is that
the identity state is not a completely ``good'' state of OSFT star
algebra, since it has several ``anomalous'' properties, discussed for
example in \cite{Rastelli:2000iu},\cite{Kishimoto:2001de},
\cite{Schnabl:2002gg}.  The argument of \cite{Ellwood:2006ba} is that any
\ state $\psi$ which is $\Qtilde$-closed is also $\Qtilde$-exact since
Eqs. (\ref{istate}) and (\ref{astate}) imply
\be
\psi = \Qtilde \, \bigl(A\star \psi\bigr)
\ee
The question we are raising therefore is if the state $A$ is well-defined
or, more precisely, if the star product of $A$ with {\it any} string field
is well-defined.

We think this question deserves further investigation. Here, we limit
ourselves to observe that the assumption of the existence of the identity
leads to consequences that look quite dramatic from the point of view of
the gauge-fixed second quantized theory.  We explained that the absence of
physical states is a statement regarding the null vectors of the
gauge-fixed kinetic operators $\Ltilde^{\ss n}(p)$ for {\it any} $p^2$,
{\it any} ghost number $n$ and {\it any} Lorentz quantum number.  Now,
assuming that the identity state does exist, one could consider, in
analogy with (\ref{astate}), the following state
\be
{\tilde A} = {1\over \Ltilde} b_0\mathcal{I}
\label{atildestate}
\ee
Then 
\be
\Qtilde\, {\widetilde A} = \mathcal{I}- P_ {\Vtilde{-1}}\mathcal{I}
\label{atildetrivial}
\ee
where $P_{\Vtilde{-1}}$ is the projector on the subspace $\Vtilde{-1}$ of
states that are in the kernel of $\Ltilde$, have momentum $p=0$, ghost
number -1 and are twist-parity even Lorentz scalars.  Therefore if
$\Vtilde{-1}$ is empty, ${\widetilde A}$ trivializes the identity.  In
other words, the existence of the identity implies that a property of the
propagators at $p^2=0$ in some definite Lorentz and ghost sector --- the
vanishing of $\Vtilde{-1}$ --- would determine the behaviour of the
propagators for all $p^2$ and all quantum numbers.  This seems a very (and
maybe too) strong statement and, we feel, suggests caution when
manipulating the identity.

Note that the vanishing of $\Vtilde{-1}$ is a question that {\it can}
reliably addressed in the LT expansion, since it involves a sector with
vanishing momentum, definite twist parity, Lorentz and ghost quantum
numbers. To test the emptiness of $\Vtilde{-1}$ it is enough to compute
the determinants of the twist-even scalar kinetic operators at zero
momentum and ghost number -1 and -2. In fact, this is a very particular
case of the computation we performed in this paper (and in
\cite{Giusto:2003wc}, for that matter). For $p=0$ our level 10
approximation should by all means be reliable: the same sector (but with
ghost number 0) is the one where the same approximation turned out to
capture quite accurately the properties of the stable classical vacuum
solution.  Since we have not detected any zeros of the determinants of the
twist-even scalar kinetic operators (for both ghost numbers -1 and -2) at
$p^2=0$, we can confidently assert that $\Vtilde{-1}$ is empty.

In this sense therefore the numerical computations presented both in this
paper and in \cite{Giusto:2003wc} are coherent with what proven in
\cite{Ellwood:2006ba}. If the expression in Eqs. (\ref{atildestate})
(analogous to Eqs.(\ref{astate})) defined a ``good'' state, our numerical
computations, when restricted to $p=0$, ghost number -1, and to the
twist-even scalar sector, could be interpreted as a reliable numerical
proof of the triviality of the identity. The extension of the same
analysis to non-vanishing $p^2$ appears to contradict this conclusion,
however: in summary, this might signals either a failure of LT when
extended to $p^2 < 0$ or the formal character of expressions like
(\ref{atildestate}) and (\ref{astate}).

In order to elucidate this question it should be helpful to extend our
computations to gauges different than the Siegel gauge. This would allow
testing the gauge invariant meaning of the propagator double poles. We
leave this to future work.

We add a final comment. In \cite{Kishimoto:2002xi} the spectrum of OSFT
around so-called universal solutions was studied analytically. It was
found there that for such background while $\H{0}$ vanishes, BRS
cohomologies with ghost number -1 and -2 are not empty.  Although this
result is intriguingly reminiscent of ours, it also differs from it in
various respects.  Cohomologies at non-standard ghost numbers around the
universal solutions are isomorphic to perturbative, ghost number zero,
cohomologies. In particular they exist for $-p^2=m^2=-2,0,2, ..$. We do
not find cohomologies for $m^2=-2,0$. Moreover the cohomologies that we do
find at $m^2=2,4,6$ do not have the same quantum numbers as the
perturbative ones.  More work is needed to understand the relation, if
there is one, between cohomologies around the stable classical solution
and around universal solutions.

The rest of this paper is organized as follows. In Section
\ref{sec:reviewgaugefixing} we briefly review, for self-containedness, the
gauge-fixing procedure of OSFT in the classical stable vacuum. In Section
\ref{sec:relabscohomologies} we derive, relaxing the additional technical
assumptions of \cite{Giusto:2003wc}, the long exact cohomology sequence
(\ref{nonpertbottsequenceintro}). In Section \ref{sec:numerical} we report
the result of our numerical level (10,30) computation. In Section
\ref{sec:qaction} we work out the unique action (\ref{mhataction}) of
$\Mhat$ on the null vectors of the gauge-fixed kinetic operators, which
should correspond, in the exact theory, to the approximately degenerate
propagators poles found in Section \ref{sec:numerical}.

\sectiono{Gauge-fixed Open String Field Action}
\label{sec:reviewgaugefixing}

The open string field theory (OSFT) action around the tachyonic background
writes
\be
\Gtilde[\Psi] = {1\over 2} \bigl( \Psi, \Qtilde\, \Psi\bigr) +
{1\over 3}\bigl(\Psi, \Psi\star\Psi\bigr)
\label{action}
\ee
$\Psi$ is the classical open string field, a state in the open string Fock
space of ghost number zero. $(A,B)$ is the bilinear form between states
$A$ and $B$ of ghost numbers $g_A$ and $g_B$ respectively. $(A,B)$
vanishes unless $g_A+g_B =1$.  $\star$ is Witten's associative and
non-commutative open string product.  $\Qtilde$ is the BRS operator around
the non-perturbative vacuum $\phi$
\be
\Qtilde\Psi \equiv Q\,\Psi + \bigl[\phi\,\mathop{,}^{\star} \Psi\bigr]
\ee
where
\be
\bigl[ A\, {\mathop{,}^{\star}} B\bigr]\equiv A\star B -
(-)^{(g_{\sss A}+1)\, (g_{\sss B}+1)}
B\star A
\ee
and $Q$ is the perturbative BRS operator, which is (anti)symmetric with
respect to the bilinear inner product $(\,\cdot\,,\,\cdot\,)$ based on BPZ
conjugation.  $\phi$ is the solution of the classical equation of motion
\be
Q\, \phi + \phi\star\phi =0
\label{flatness}
\ee
that represents the tachyonic vacuum.  The flatness equation
(\ref{flatness}), together with the associativity of the $\star$-product,
ensures the nilpotency of $\Qtilde$. $\Qtilde$ is (anti)symmetric with
respect to the product $(\,\cdot\,,\,\cdot\,)$ thanks to the property
\be
(A,\,\phi\star B)=  (A\star\phi, B)
\ee
The action (\ref{action}) is thus invariant under the following gauge
transformations
\be
\delta\, \Psi = \Qtilde\, C + \bigl[\Psi\,\mathop{,}^{\star} C\bigr]
\label{gaugeinvariance}
\ee
where $C$ is a ghost number -1 gauge parameter.

CFT ghost number $g$ provides a grading for string fields: ``matter''
string field have $g=0$.  It is useful to introduce another grading, the
second quantized string field ghost number, that we will denote by
$n_{\sss sft}$. Matter fields have $n_{\sss sft} =0$, by definition.
Fields with second quantized ghost number $n_{\sss sft}=n$ and CFT ghost
number $g$ will be denoted with $\Psi^{\ss{n}}_{\sss g}$.

The gauge invariance (\ref{gaugeinvariance}) of the classical OSFT action 
translates into the second quantized BRS symmetry
\be
\brs\, \Psi^{\ss0}_{\sss 0} = \Qtilde\, \Psi^{\ss{1}}_{\sss -1} + 
\bigl[\Psi^{\ss0}_{\sss 0}\,\mathop{,}^{\star} \Psi^{\ss1}_{\sss -1}\bigr]
\label{brsinvariance}
\ee
where $\Psi^{\ss1}_{\sss -1}$ is the ghost string field of first generation.

We will gauge-fix the invariance (\ref{brsinvariance}) by going to  
Siegel gauge:
\be
b_0 \, \Psi^{\ss0}_{\sss 0} = 0
\label{siegelgauge}
\ee
Gauge-fixing the OSFT action requires an infinite number of ghost field
generations \cite{bochicchiothorn}. We will adopt the Siegel gauge for all
higher-generation ghost string fields:
\be
b_0\, \Psi^\ss{n}_{\sss -n} =0
\ee
For any field $\Psi^\ss{n}_{\sss m}$ one can write the decomposition
\be 
\Psi^{\ss{n}}_{\sss m} = \phi^{\ss{n}}_{\sss m} + c_0\, \phi^\ss{n}_{\sss
m-1}
\ee
where $\phi^\ss{n}_{\sss m}$ and $\phi^\ss{n}_{\sss m-1}$ 
are fields that do not contain $c_0$:
\be
b_0\, \phi^\ss{n}_{\sss m}\, = 0\qquad\forall\,\, m,\, n  
\ee
The corresponding, completely gauge-fixed, quadratic action is
\bea
&& \Gtilde_{g.f.}^{\ss2} = {1\over 2} 
\bigl( \phi^{\ss0}_{\sss 0}, c_0\, \Ltilde\, \phi^{\ss0}_{\sss 0}
\bigr)+ \sum_{n=1}^\infty
\bigl(\phi^\ss{-n}_{\sss n}, c_0\, \Ltilde\, \phi^\ss{n}_{\sss -n}
\bigr)
\label{fpactioninfinity}
\eea 
where
\be
\Ltilde \equiv \{ \Qtilde, b_0\}
\label{kinetictilde}
\ee
Thus the gauge-fixed OSFT action depends on fields $\phi^\ss{-n}_{\sss
n}\equiv \varphi_{\sss n}$ which are $b_0$-invariant states of the first
quantized Fock space with CFT ghost number $n$ and second quantized ghost
number $-n$. We will denote this state space with $\O{n}$.

It is convenient to define the following {\it non-degenerate} bilinear
form $\langle \cdot,\cdot\rangle$ on $\O{-n} \times \O{n}$
\be
\langle \cdot, \cdot\rangle \equiv (\,\cdot\,, c_0\, \cdot\,)
\ee
From the definition (\ref{kinetictilde}) of $\Ltilde$ and from the Jacobi
identity one obtains:
\be
[\Ltilde, c_0] = [\{\Qtilde,b_0\},c_0] = [b_0, \{\Qtilde,c_0\}] = 
[b_0, \Dtilde]
\label{iacobi}
\ee
where $\Dtilde \equiv \{\Qtilde, c_0\}$.  This ensures that $\Ltilde$ is
an operator on $\O{n}$ which is symmetric with respect the bilinear form
$\langle\, \cdot, \cdot\,\rangle$:
\be
\langle \varphi_{\sss n}, \Ltilde\, \varphi_{\sss -n}\rangle =
\langle\Ltilde\,  \varphi_{\sss n} , \varphi_{\sss -n}\rangle
\label{lhermitian}
\ee
In conclusion the quadratic part of the gauge-fixed OSFT action at the
tachyonic background writes as
\be
\Gtilde_{g.f.}^{\ss2} =
{1\over 2} 
\langle \varphi_{\sss 0}, \Ltilde\, \varphi_{\sss 0}\rangle
+ \sum_{n=1}^\infty
\langle\varphi_{\sss n}, \Ltilde\, \varphi_{\sss -n}\rangle
\label{gaugefixedquadraticbis}
\ee

\sectiono{Relative and Absolute Cohomologies }
\label{sec:relabscohomologies}
Let $\F{n}$ be the space of states of CFT ghost number $n$. Let us denote
by $\H{n}(\Qtilde)$ the $\Qtilde$-cohomologies on $\F{n}$. We will refer
to $\H{n}(\Qtilde)$ as the {\it absolute} BRS state cohomologies.  As we
recalled in the Introduction, the number of physical states of open string
theory is given by the dimension of the $\H0(\Qtilde)$ cohomology.

One way to compute $\H{n}(\Qtilde)$ is based on the preliminary
computation of a different kind of $\Qtilde$-cohomologies --- the {\it
relative} cohomologies.  Let $\Wtilde{n}$ be the subspace of $\F{n}$ of
states $\phi_n$ of ghost number $n$ which are both $b_0$ and $\Ltilde$
invariant:
\be
\phi_n\in \Wtilde{n}\,  
\mathrel{\mathop{\iff}^{\rm def}}\, b_0\,\phi_n =\Ltilde\,\phi_n =0
\ee
The {\it relative} $\Qtilde$-cohomology of ghost number $n$ is given by
the $\Qtilde$-closed states $\phi_n\in \Wtilde{n}$
\be
\Qtilde\, \phi_n =0
\ee
modulo the states which are in the $\Qtilde$ image of $\Wtilde{n-1}$
\be
\phi_{\sss n} \sim \phi^\prime_{\sss n} =
\phi_{\sss n} + \Qtilde\, \phi_{\sss n-1}
\ee
where $\phi_{\sss n-1}\in \Wtilde{n-1}$. Such a definition is
consistent since 
\be
\{\Qtilde, b_0\} = \Ltilde
\label{dueltilde}
\ee
We will denote the relative cohomologies of $\Qtilde$ by $\htilde{n}$.

Let us decompose $\Qtilde$ in the  $b_0, c_0$ algebra:
\be
\Qtilde = c_0\, \Lhat + b_0\,\Dhat + \Mhat  + c_0\, b_0\, \Zhat
\label{nonpertdecomposition}
\ee
where $\Lhat$, $\Dhat$, $\Mhat$ and $\Zhat$ are independent of $c_0$ and
$b_0$.  The crucial difference between the decomposition
(\ref{nonpertdecomposition}) of the non-perturbative $\Qtilde$ and its
perturbative analogue is the term proportional to $c_0\, b_0$, which is
absent in the perturbative case. Note that
\be
\Ltilde \equiv \{\Qtilde, b_0\} = \Lhat + b_0\, \Zhat 
\qquad \Dtilde \equiv \{\Qtilde, c_0\} = \Dhat - c_0\,\Zhat
\label{zed}
\ee
and therefore $[\Ltilde, c_0] = [b_0,\Dtilde]= -\Zhat$, in agreement with
the Jacobi identity (\ref{iacobi}).  The first equation of (\ref{zed})
implies that $\Wtilde{n}$ is the kernel of $\Lhat$ on $\O{n}$ (i.e. the
space of $b_0$-invariant states of ghost number $n$):
\be
\Wtilde{n} = \ker\Lhat^{(n)}
\ee
The nilpotency of $\Qtilde$ are equivalent to the following equations
\bea
&&\Mhat^2 + \Dhat\,\Lhat =0 \qquad \{\Mhat, \Zhat\} + \Zhat^2 = [\Dhat,
\,\Lhat]
\nonumber\\
&&\Lhat\, \Mhat - (\Mhat +\Zhat)\, \Lhat =0\qquad
\Mhat\,\Dhat -\Dhat\, (\Mhat +\Zhat) =0
\label{nilnonpert}
\eea
These equations show that the
$b_0$-relative cohomology $\htilde{n}$ is the cohomology of the
operator $\Mhat$ on $\Wtilde{n}$:
\be
\htilde{n} = \H{n} \bigl(\Mhat, \Wtilde{n} \bigr)
\ee
Indeed, the first of the equations
(\ref{nilnonpert}) says that $\Mhat^2=0$ on $\Wtilde{n}$
and the third of the equations (\ref{nilnonpert}) 
guarantees that $\Mhat: \Wtilde{n}\to\Wtilde{n+1}$. 

Let us denote by 
$\Vtilde{n}$  the kernel of $\Ltilde$ on $\F{n}$:
\be
\Vtilde{n} = \ker\Ltilde^{(n)}
\ee
Thanks to Eq. (\ref{dueltilde}), the cohomology of $\Qtilde$ on $F_n$ is
identical to the cohomology of $\Qtilde$ on $\Vtilde{n}$.

Now we come to the main point. We want to describe the cohomology of
$\Qtilde$ on $\Vtilde{n}$ in terms of cohomologies defined on the
$\Wtilde{n}$'s, the gauge-fixed ($b_0$-invariant) spaces. There exists two
natural maps between these spaces: the immersion map $\imath$
\be
\imath\, : \Wtilde{n} \rightarrow \Vtilde{n}\qquad \imath(\phi_n) =\phi_n
\label{immersion}
\ee
and the  projection  $\pi$:
\be
\pi\, : \Vtilde{n} \rightarrow \Wtilde{n-1}\qquad
\pi(\phi_n+c_0\phi_{n-1}) =\phi_{n-1}
\label{projection}
\ee
The problem is that, although $\imath$ is injective, the projection $\pi$
is not in general surjective --- if $\Zhat$ is not vanishing. For this
reason we introduce the image of $\Vtilde{n}$ by the map $\pi$ and denote
it by $\Wcheck{n-1}$:
\be
\phi_{n-1}\in \Wcheck{n-1}\,  
\mathrel{\mathop{\iff}^{\rm def}}\, \phi_{n-1}\in\Wtilde{n-1},\,\, 
\Zhat\, \phi_{n-1} = \Lhat\, \phi_{n}, \,\, \phi_{n}\in \O{n} 
\label{checkdef}
\ee
$\Wcheck{n}$ is in general a subspace of $\Wtilde{n}$ which reduces to the
latter when $\Zhat$ vanishes.

Therefore, by construction, the following is an exact short sequence
\be
0\; \longrightarrow \Wtilde{n}\; {\mathop{\longrightarrow}^{\imath }}\;
\Vtilde{n}\;{\mathop{\longrightarrow}^{\pi}}\;\Wcheck{n-1}\; 
\longrightarrow\; 0
\label{bottsequence1}
\ee

Moreover $\Qtilde\, \imath = \imath\, \Mhat$ and $\Mhat\, \pi = -\pi\,
\Qtilde$. Also, it is easily verified that
\be
\Mhat :\Wcheck{n} \to \Wcheck{n+1}
\label{wcheckinvariance}
\ee
by virtue of the nilpotency relations (\ref{nilnonpert}).

In conclusion, the following diagram is (anti)-commutative

\be
\def\normalbaselines{\baselineskip20pt \lineskip3pt\lineskiplimit3pt}
\def\mapright#1{\smash{\mathop{\longrightarrow}\limits^{#1}}}
\def\mapdown#1{\Big\downarrow\rlap{$\vcenter{\hbox{$\scriptstyle#1$}}$}}
\matrix{
0&\mapright{}&\Wtilde{n}&\mapright\imath&\Vtilde{n}
&\mapright\pi&\Wcheck{n-1}&\mapright{}&0\cr
&&\mapdown\Mhat&&\mapdown\Qtilde&&\mapdown\Mhat\cr
0&\mapright{}&\Wtilde{n+1}&\mapright\imath&\Vtilde{n+1}
&\mapright\pi&\Wcheck{n}&\mapright{}&0\cr
}
\label{commutativediagram}
\ee

\noindent From this diagram, one obtains (see, for example, \cite{bott}),
along the usual lines, the following {\it exact} long sequence of
$\Qtilde$-cohomologies
\be
\cdots\; {\mathop{\longrightarrow}^{\Dcaltilde}}\; \htilde{n}\;
{\mathop{\longrightarrow}^{\imath }}\;\H{n}(\Qtilde)\;
{\mathop{\longrightarrow}^{\pi}}
\;\hcheck{n-1}\;{\mathop{\longrightarrow}^{\Dcaltilde}}\;\htilde{n+1}\;
{\mathop{\longrightarrow}^{\imath}}\;\cdots
\label{nonpertbottsequence}
\ee

In this exact sequence a new kind of relative cohomology appears,
$\hcheck{n}$, to which we will refer as the {\it check relative}
cohomology. This is defined as the cohomology of $\Mhat$ on $\Wcheck{n}$:
\be
\hcheck{n} = \H{n} \bigl(\Mhat, \Wcheck{n} \bigr)
\ee

The map $\Dcaltilde$ is known in homology theory as the ``connecting map''
and it is defined as follows. Let $v_{n-1}$ be an element of
$\Wcheck{n-1}$. Thus, there exists $\phi_n\in \F{n}$ such that
\be
\Zhat\, v_{n-1} = \Lhat\, \phi_n
\ee
Therefore $\phi_{n}+c_0\, v_{n-1}\in \Vtilde{n}$ and 
\be
\pi\bigl(\phi_{n}+c_0\, v_{n-1}\bigr)= v_{n-1}
 \ee
We define
\be
 \Dcaltilde (v_{n-1}) \equiv \Dhat\, v_{n-1} +
\Mhat\,\phi_{n}
\label{dcaltilde}
\ee
The commutativity of the diagram (\ref{commutativediagram}) and the
nilpotency relations (\ref{nilnonpert}) ensure that $\Dcaltilde$ descends
to a cohomology map.

Indeed, suppose $v_{n-1}\in \Wcheck{n-1}$ is in the kernel of
$\Mhat$. Then
\bea
&& \Mhat \,\Dcaltilde (v_{n-1}) = \Mhat\, \Dhat\, v_{n-1} +
\Mhat^2\,\phi_{n}= \Dhat\, (\Mhat+ \Zhat) \,v_{n-1}-
\Dhat\,\Lhat\,\phi_n=\nonumber\\
&&\quad = \Dhat\, (\Mhat+ \Zhat) \,v_{n-1}- \Dhat\,\Zhat\,
v_{n-1}=0\nonumber
\eea
and
\bea
&& \Lhat \,\Dcaltilde (v_{n-1}) = \Lhat\, \Dhat\, v_{n-1} + (\Mhat+
\Zhat)\, \Zhat \,v_{n-1}= \Dhat\, \Lhat \,v_{n-1}-
\Zhat\,\Mhat\,v_{n-1}=0\nonumber
\eea
Therefore $\Dcaltilde$ maps the kernel of $\Mhat$ on $\Wcheck{n-1}$ to
kernel of $\Mhat$ on $\Wtilde{n+1}$.

Suppose now that $v_{n-1}$ is trivial in check cohomology:
\be
v_{n-1} = \Mhat\, v_{n-2}\qquad \Lhat\, v_{n-2}=0\qquad \Zhat\, v_{n-2}
=\Lhat\,\phi_{n-1}\nonumber
\ee
Hence
\bea
&&\Dcaltilde (v_{n-1}) =\Dhat\,\Mhat v_{n-2} +
\Mhat\,\phi_{n}= \Mhat\,\Dhat\, v_{n-2} -\Dhat\,\Zhat\, v_{n-2}+
\Mhat\,\phi_{n}=\nonumber\\
&&\quad =\Mhat\,\Dhat\, v_{n-2} -\Dhat\,\Lhat\phi_{n-1}+
\Mhat\,\phi_{n}=\Mhat\bigl(\Dhat\, v_{n-2} +\Mhat\,\phi_{n-1}+
\phi_{n}\bigr)\nonumber
\eea
Moreover
\bea
&&\Lhat\, \bigl(\Dhat\, v_{n-2} +\Mhat\,\phi_{n-1}+ \phi_{n}\bigr)=
\Lhat\, \Dhat\, v_{n-2}+ (\Mhat+\Zhat)\,
\Zhat\,v_{n-2}+\Zhat\,v_{n-1}=\nonumber\\ 
&&\quad =\Lhat\, \Dhat\, v_{n-2}+ (\Mhat+\Zhat)\, \Zhat\,v_{n-2}+
\Zhat\,v_{n-1}=\nonumber\\ 
&& \quad= \Dhat\, \Lhat\, v_{n-2} -\Zhat\, \Mhat\, v_{n-2}+\Zhat\,\Mhat\,
v_{n-2}=0\nonumber
\eea
Thus $\Dcaltilde$ maps trivial states to trivial states.  Hence,
Eq. (\ref{dcaltilde}) defines a map between $\hcheck{n-1}$ and
$\hcheck{n+1}$.

We have observed that in the perturbative case $\Zhat=0$ and therefore 
$\htilde{n}=\hcheck{n}$. In this case, therefore, the sequence
(\ref{nonpertbottsequence}) allows determining the absolute cohomologies
by means of the relative ones. On the other hand, if $\Zhat\not=0$, the
knowledge of both $\htilde{n}$ and $\hcheck{n}$ is needed, in general, for
the computation of the absolute cohomologies by means of the exact
sequence.  One can however derive few general relations connecting tilde
and check relative cohomologies.

One such relations is the following: define the tilde relative index, 
\be
\mathrm{index}\, \tilde{\mathrm{h}}= \sum _{n} (-1) ^n \dim \htilde{n}
\ee
and the check relative index
\be
\mathrm{index}\, \check{\mathrm{h}}= \sum _{n} (-1) ^n \dim \hcheck{n}
\ee
Then, the duality between absolute cohomologies,
\be
\H{n}(\Qtilde)\approx \H{1-n}(\Qtilde)
\ee
together with the sequence (\ref{nonpertbottsequence}) leads to the 
identity of the relative cohomology indices:
\be
\mathrm{index} \,\tilde{\mathrm{h}}=\mathrm{index}\, \check{\mathrm{h}}
\label{relativeindices}
\ee

Further relations between  tilde and check cohomologies 
are somewhat obvious and yet useful inequalities which rest on
mathematical properties of the operators 
$\Lhat^{(n)}$ that one can assume on physical grounds. Let us list 
such properties:

a) For the Siegel gauge to be a ``good'' gauge, the kernels of
$\Lhat^{(n)}$ --- i.e. $\Wtilde{n}$ --- must vanish for $p^2$ {\it
generic}.  This is equivalent to the requirement that propagators be
well-defined after gauge-fixing.

b) It is also physically reasonable to assume that the dimensions of the
kernels $\Wtilde{n}$ --- at a given discrete value of $p^2$ for which they
are not empty --- remain {\it finite}.  This amounts to say that we expect
a finite number of fields of a given mass.

c) Last, we should assume that at a given value of $p^2$ there is only a
finite number of $\Wtilde{n}$ with different ghost number $n$ that are
non-empty: in other words, for a given $p^2$, there exists a maximal ghost
number $g>0 $ such that $\Wtilde{n}=0$ for $|n|> g$. This assumption is
essential to give a mathematical precise meaning to the Fadeev-Popov index
(\ref{fpindex}) that counts the number of physical states. More generally,
this assumption gives mathematical sense to the BRS gauge-fixing
construction for OSFT  which, as we have seen, involves an
infinite number of ghost fields generations.

All these three conditions are obviously verified in the level truncated
theory, for fixed level $L$. The validity of LT as a computational scheme
of OSFT is based on the assumption that these properties are ``stable'' as
$L\to\infty$.  This means that for a given interval of $p^2$ there should
exist a level ${\bar L}$ such that for levels $L> {\bar L}$ the dimensions
of the $\Wtilde{n}$ do not jump even if the values of $p^2$ at which
non-trivial $\Wtilde{n}$'s appear move a bit.  To state it a little more
precisely: given $p^2$, if $\dim \Wtilde{n}(p_L)\not= 0$ for $p_L^2 > p^2
$ and $L>{\bar L}$, then for any $L'>L$ there should exist a $p_{L'}^2$
for which $\dim \Wtilde{n}(p_{L'})=\dim \Wtilde{n}(p_L)$ and
$|p_{L'}^2-p^2_L| \to 0$ as both $L$ and $L'$ go to infinity.

Let us remark that we are {\it not} assuming uniform convergence on the
$p^2$ axis: ${\bar L}$ may well depend on $p^2$ and, indeed, our numerical
computations suggest that it grows linearly as $p^2 \to -\infty$.
   
We have no formal proof of this ``stability'' property of level
truncation, although our numerical computation are consistent with it.  On
the other hand, if level truncation did not enjoy this property its use in
OSFT would have in general no justification, putting aside the specific
problem we are considering.

Let us now come back to the inequalities between dimensions of
relative cohomologies that one can prove assuming a)-c) in the exact 
theory. Let $g>0$ be the {\it maximal} ghost number, such that 
$\Wtilde{n}=0$ for $|n|> g$, as specified in c).  The image of $\Mhat$ 
in $\Wtilde{-g}$ vanishes, since $\Wtilde{-g-1}=0$.  
Therefore $\htilde{-g}$ reduces to the kernel of $\Mhat$ on 
$\Wtilde{-g}$ while $\hcheck{-g}$ is the kernel of
$\Mhat$ restricted to $\Wcheck{-g}\subset \Wtilde{-g}$.  Therefore
\be
\dim  \hcheck{-g} \leq \dim \htilde{-g}
\label{nless}
\ee
An analogous inequality is derived as follows. The kernel of 
$\Mhat$ at ghost number $g$ consists of the whole
$\Wtilde{g}$, since $\Wtilde{g+1}=0$.  Given any vector $v_g$ in
$\Wtilde{g}$ we can decompose it as follows
\be
\Zhat v_{g} = \Lhat\, \phi_3 + v_{g+1}^*
\label{gmax}
\ee
where $\phi_3\in\Omega_3$ and $v_g^*$ is an element of the cokernel
$\mathrm{coker} \Lhat^{(g+1)}$ of $\Lhat$ on $\Omega_3$
\be
v_g^*\in \mathrm{coker} \Lhat^{(g+1)}\equiv \Omega_3/
\mathrm{img}\,\Lhat^{(g+1)}
\ee
By hypothesis $\Wtilde{g+1}=\ker \Lhat^{(g+1)}=0$: therefore 
\be
 \dim \mathrm{coker} \Lhat^{(g+1)}=\dim \ker \Lhat^{(g+1)}= 0
 \ee
In other words,  $ v_{g+1}^*=0$ in the equation (\ref{gmax}) above and 
 \be
 \Wtilde{g}=\Wcheck{g}= \ker \Mhat^{(g)}
 \label{wcheckmax}
 \ee  
On the other hand, $\Wcheck{g-1}\subset\Wtilde{g-1}$ and thus the image of
$\Wcheck{g-1}$ via $\Mhat$ is contained in the image of $\Wtilde{g-1}$. We
conclude that
\be
\dim  \hcheck{g} \geq \dim \htilde{g}
\label{ngreater}
\ee

\sectiono{The numerical computation}
\label{sec:numerical}
The field spaces $\O{n}$ can be decomposed as direct sum of spaces with
fixed space-time momentum $p^\mu$, $\mu=0,1,\ldots,25$: \be \O{n} =
\oplus_{p}\, \O{n}(p) \ee Because of translation invariance the kinetic
operator $\Ltilde$ is diagonal with respect to this decomposition. For
each space $\O{n}(p)$ choose a basis $\{e^{\sss (n)}_{i_n}(p)\}$. Let us
denote by $\Ltilde^\ss{n}(p)$ the matrix representing in this basis the
operator $\Ltilde$ acting on $\O{n}(p)$.  Let $G^{\sss (n)}(p)$ be the
square matrix whose elements are given by
\be
(G^{\sss (n)}(p))_{i_n\,j_n} = \langle e^{\sss (-n)}_{i_n}(p)\, ,\, 
e^{\sss (n)}_{j_n}(p)\rangle
\ee
For $n>0$ the symmetric square matrix that 
specifies the kinetic operator for the fields 
$(\varphi_{\sss -n},\varphi_{\sss n})$ is
\be
C^{\sss (-n)}(p) \equiv {1\over 2}
\left(\matrix{0& G^{\sss (n)}(p)\, \Ltilde^{\ss{n}}(p)\cr
G^{\sss (-n)}(p)\,\Ltilde^{\ss{-n}}(p) & 0\cr}\right)
\label{covariancen}
\ee
For the ``matter'' string field $\varphi_{\sss 0}$ the kinetic
quadratic form is instead
\be
C^{\sss (0)} \equiv  G^{\sss (0)}(p)\,\Ltilde^{\ss{0}}(p)
\label{covariancezero}
\ee
The determinants of the kinetic operators
\be
\Delta^\ss{n}(p^2) \equiv \det \Ltilde^\ss{n}(p)
\ee
 are functions of $p^2$.  The  zeros of such determinants encode
 the information about physical states of OSFT.   Suppose that 
\be
\Delta^\ss{n}(p^2) = \Delta^\ss{-n}(p^2)  
= a_n\, (p^2+m^2)^{d_n} (1+ O(p^2+m^2))
\ee
where the first equality is a consequence of the symmetry property
(\ref{lhermitian}) of $\Ltilde$. 
Then, the number of physical states of mass $m$ is given by the
index:
\be
I_{FP}(m) = d_0 - 2\, d_1 + 2\, d_2 +\cdots = \sum_{n=-\infty}^\infty 
(-1)^n\, d_n
\label{fpindex}
\ee
This is so since the ghost and anti-ghost pairs $(\varphi_{\sss -n},
\varphi_{\sss n})$ are complex fields of Grassmanian parity $(-1)^n$.  The
numbers $d_n$ are in general gauge-dependent --- in our case they capture
properties of the $b_0$-invariant spaces $\O{n}$. The index $I_{FP}(m)$ is
gauge-invariant and coincides with the dimension of the cohomology
$\H0(\Qtilde)$ of $\Qtilde$ on the total space of (non-$b_0$-invariant)
states of ghost number 0.  In a physically sensible theory $I_{FP}(m)$
must be non-negative.  Sen's conjecture is that $I_{FP}$ vanishes for all
$m$.

Typically, in the exact (not level truncated) theory,
$\Delta^\ss{n}(p^2)$'s with different ghost numbers $n$ vanish at the same
value of $p^2$, as a consequence of BRS invariance. Indeed $[\Qtilde,
\Ltilde] = 0$; so, if $\Delta^\ss{n}(p^2)$ vanishes for some $p^2=-m^2$,
then there exists a $\varphi_{\sss n}$ such that
\be
\Ltilde\,\varphi_{\sss n}= 0 = b_0 (\Qtilde\,\varphi_{\sss n})
\ee
Therefore 
\be
\Ltilde(\varphi_{\sss n+1})= 0 = b_0\,\varphi_{\sss n+1}
\ee
where $\varphi_{n+1}=\Qtilde\,\varphi_{\sss n}$. If $\varphi_{n+1}$ does
not vanish, $\Delta^\ss{n+1}(-m^2)=0$. Thus physical states of mass $m^2$
are associated to a multiplet of determinants $\Delta^\ss{n}(p^2)$ with
different $n$'s that vanish simultaneously at $p^2=-m^2$.

Since level truncation breaks BRS invariance we expect that the zeros of
the determinants in the same multiplet, when evaluated at finite $L$,
would be only approximately coincident. Thus using the index formula
(\ref{fpindex}) to compute the number of physical states is meaningful
when the splitting between approximately coincident determinant zeros is
significantly smaller than the distance between the masses of different
multiplets.

In the theory truncated at level $L$, the operators $\Ltilde^\ss{n}(p)$
reduce to finite dimensional matrices; moreover for a given $L$, the
$\Ltilde^\ss{n}(p)$ vanish identically for $n$ greater than a certain
$n_L$ which depends on the level\footnote{$n_L$ is the greatest integer
which satisfies the inequality $n_L(n_L+1)/2 \le L$.}.  We evaluated the
LT matrices $\Ltilde^\ss{n}(p)$ on both ${\O{n}}^{\sss \!\!\!scalar}(p)$
and ${\O{n}}^{\sss \!\!\!vector}(p)$, the subspaces of $\O{n}(p)$
containing the states which are either scalars or vectors with respect to
space-time Lorentz symmetry.

The computation is simplified by noting that the non-perturbative
$\Qtilde$ commutes with the twist parity operator $(-1)^{\hat{\rm N}}$.
Therefore the kinetic operators decompose as follows 
\be
\Ltilde^\ss{n}(p) = \Ltilde^\ss{n,+}(p)\oplus\Ltilde^\ss{n,-}(p)
\ee
where $\Ltilde^\ss{n,\pm}(p)$ are the kinetic operators acting on the
subspaces $\O{n}^\ss{\pm}(p)$ of $\O{n}(p)$ with twist parity $\pm$.

Another symmetry of $\Ltilde$ is the $SU(1,1)$ symmetry generated by:
\bea
&&J_+ = \{Q,c_0\} =\sum_{n=1}^\infty n\, c_{-n} c_n \qquad J_- =
\sum_{n=1}^\infty {1\over n}\, b_{-n} b_n\nonumber\\
&&J_3 = {1\over 2}\,
\sum_{n=1}^\infty (c_{-n} b_n - b_{-n} c_n)
\label{sutwo}
\eea
$J_\pm$ and $J_3$ are derivatives of the $\star$-product 
\cite{Zwiebach:2000vc}. They obviously
commute both with $b_0$ and the perturbative $L_0$ and hence they
are a symmetry of the OSFT equations of motion in the Siegel gauge:
\be
L_0\, \phi + b_0(\phi\star\phi)=0
\ee
The tachyon solution turns out to be a {\it singlet} of the $SU(1,1)$
algebra: it follows that $J_\pm$ and $J_3$ commute with $\Ltilde$ since
\be
\Ltilde =L_0 + \{b_0,[\phi\,
\mathop{,}^{\!\!\!\star}\,\cdot]\}
\ee
Thus the multiplets of determinants $\Delta^\ss{n}(p^2)$ that vanish at a
given $p^2=-m^2$ organize themselves into representations of
$SU(1,1)$. The symmetry (\ref{sutwo}) is not broken by LT since its
generators commute with the level: therefore the $SU(1,1)$ symmetry of the
multiplets of vanishing determinants $\Delta^\ss{n}(-m^2)$ is exact even
at finite $L$.  Because of the $SU(1,1)$ symmetry, the Fadeev-Popov
formula for the number of physical states of mass $m$ rewrites in Siegel
gauge as follows
\be
I_{FP}(m) =  \sum_{J}
(-1)^{2\,J}(2\,J+1)\, d_J
\label{fpindexsiegel}
\ee
where the sum is over the $SU(1,1)$ spin $J$ of the representations formed
by the zeros of the determinants of the kinetic operators at $p^2=-m^2$
and $d_J$ are their associated exponents.

We computed numerically the matrices $\Ltilde^\ss{n}(p)$ as functions of
$p$ in the theory truncated at various levels $L$, from $L=4$ up to
$L=10$.\footnote{ We adopted the approximation that, in the terminology of
\cite{sz}, is of type (L, 3L). We verified that the approximation of type
$(L,2L)$ is not satisfactory for this problem.}.

For $L\le 10$ the subspaces ${\O{n}}^{\sss\!\!\!scalar}(p)$
(${\O{n}}^{\sss\!\!\!vector}(p)$) are non-empty for $|n|\le 4$ ($|n|\le
3$). The dimensions of the matrices $\Ltilde^\ss{n,+}(p)$
($\Ltilde^\ss{n,-}(p)$) for scalars and vectors at even (odd) levels are
listed in Tables~ \ref{tabellascalari} and \ref{tabellavettori}.

\begin{table}[htdp]
\caption{Number of $b_0$-invariant scalar states at up to
level 10.}
\begin{center}
\begin{tabular}{|c|c|c|c|c|c|}
\hline
Level & ghost \# 0 & ghost \# -1 & ghost \# -2 & ghost \# -3 & ghost \#
-4\\ \hline
 3 (odd)  &  9    & 6  & 1  & 0 &0 \\ \hline
4 (even) & 24   & 13  & 2  & 0  &0\\ \hline
5 (odd)  & 45   &30  & 7  & 0 & 0\\ \hline
6 (even)  & 99   & 61  & 14  & 1&0 \\ \hline
7 (odd)   & 183  & 125  & 35  & 2&0\\ \hline
8 (even)  & 363  & 240  & 68  & 7 &0\\ \hline
9 (odd)   & 655  & 458  & 145  & 15 &0\\ \hline
10 (even)   & 1216  & 841 &  272 &  36&1\\ \hline
\end{tabular}
\end{center}
\label{tabellascalari}
\end{table}%
\begin{table}[htdp]
\caption{Number of $b_0$-invariant vector states  up to 
level 10.}
\begin{center}
\begin{tabular}{|c|c|c|c|c|c|}
\hline
Level &  ghost \# 0 & ghost \# -1 & ghost \# -2  &ghost \# -3
\\ \hline
 3 (odd)  &  7   & 3  & 0  & 0  \\ \hline
 4 (even) & 16   & 9  & 1  & 0 \\ \hline
 5 (odd)  & 40   & 22  & 3 & 0 \\ \hline
 6 (even)  & 85   & 52  & 10  & 0 \\ \hline
 7 (odd)   & 184  & 113  & 24  & 1 \\ \hline
 8 (even)  & 367  & 238  & 59  & 3 \\ \hline
 9 (odd)   & 730  & 478 & 127  & 10 \\ \hline
 10 (even)   & 1385  & 936 &  272 & 25 \\ \hline
\end{tabular}
\end{center}
\label{tabellavettori}
\end{table}%

\begin{figure}[ht]
\begin{center}
\raisebox{10pt}{$\scriptstyle{\rm (a)}$}
\includegraphics*[scale=.6, clip=false]{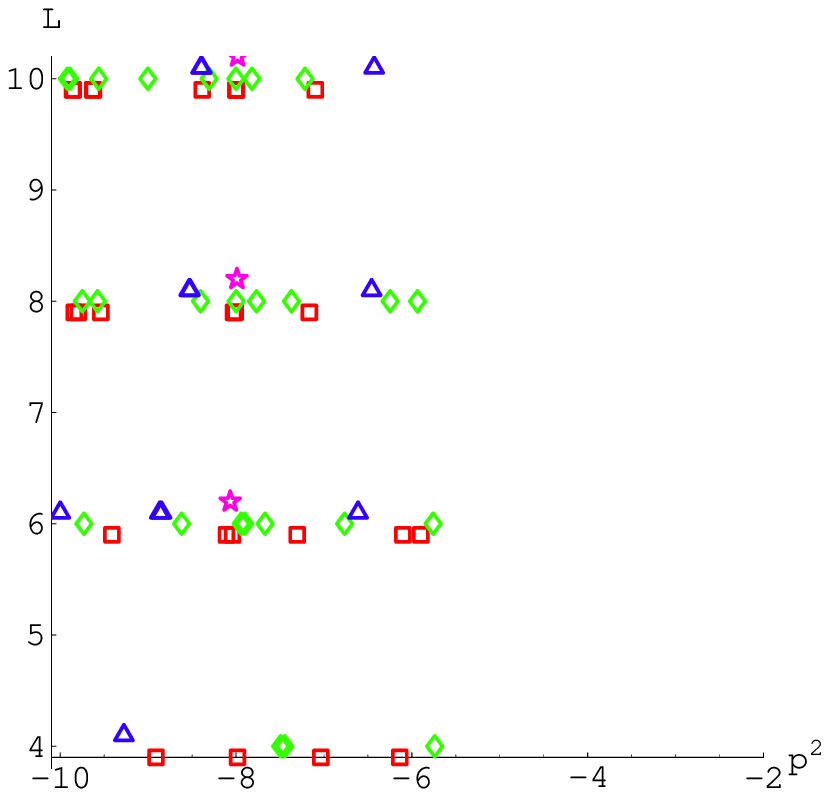}\quad
\raisebox{10pt}{$\scriptstyle{\rm (b)}$}
\includegraphics*[scale=.6, clip=false]{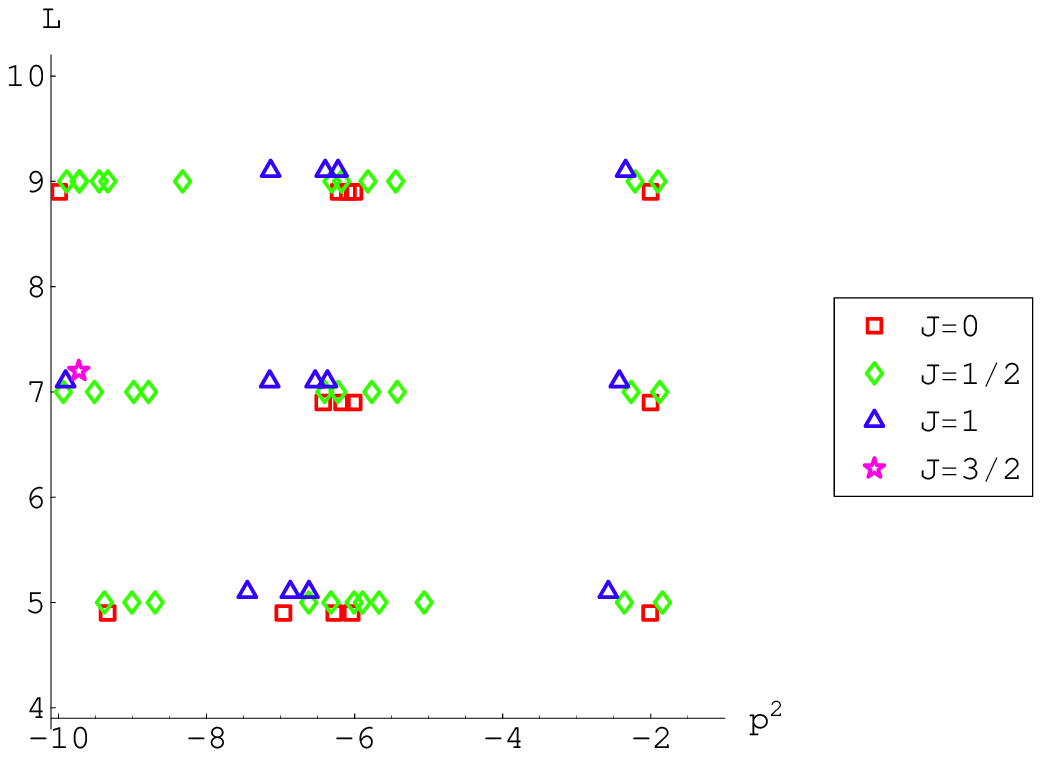}\\
\bigskip
\raisebox{10pt}{$\scriptstyle{\rm (c)}$}
\includegraphics*[scale=.6, clip=false]{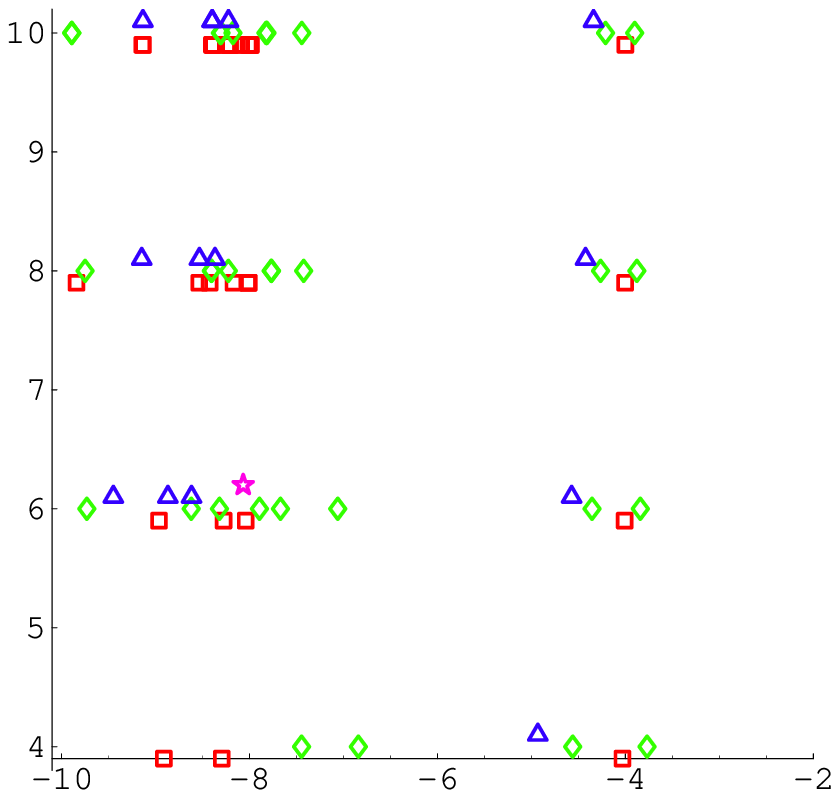}\quad
\raisebox{10pt}{$\scriptstyle{\rm (d)}$}
\includegraphics*[scale=.6, clip=false]{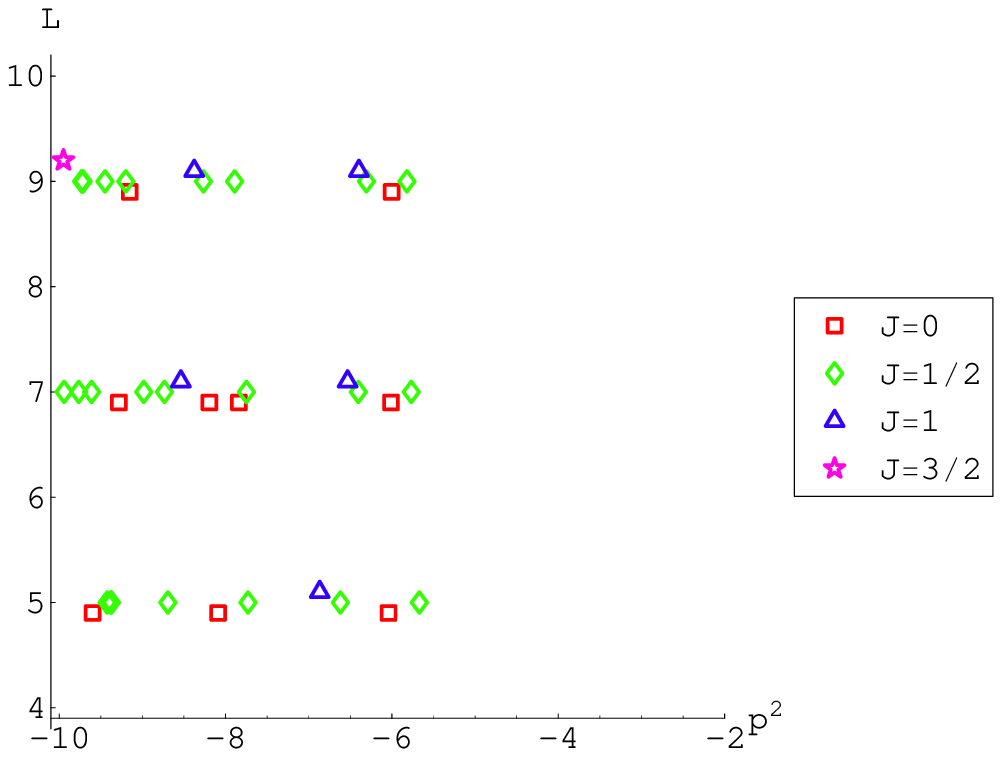}
\end{center}
\caption[x] {\footnotesize Zeros of scalar $\Delta^\ss{n}_{+}(p^2)$ (a),
scalar $\Delta^\ss{n}_{-}(p^2)$ (b), vector $\Delta^\ss{n}_{+}(p^2)$
(c),vector $\Delta^\ss{n}_{-}(p^2)$ (d) at levels $L= 4,\ldots,9$ up to
$p^2 =-10$.}
\label{f:zerospdf}
\end{figure}

We looked for zeros of  the determinants 
\be
\Delta^\ss{n}_\pm(p^2) \equiv \det \Ltilde^\ss{n,\pm}(p) 
\ee 
in the scalar and vector sector.  The zeros of the determinants for
$p^2>-10$ are plotted in Figure \ref{f:zerospdf}.

We found that the determinants have zeros only on the negative $p^2$ axis,
corresponding to physical (positive) values of $m^2=-p^2$. The first zeros
(on the negative $p^2$ axis, closest to the origin) of the scalar even
determinants are located around $p^2=-6$ (Graph (a) of Figure
\ref{f:zerospdf}). However up to level 10 they are not stable yet. Their
number keeps jumping as one increases the level: no multiple structure is
detectable up to this level. We cannot therefore draw any conclusions
about the fate of the Sen's conjecture in the even scalar sector.

\begin{figure}[ht]
\begin{center}
\raisebox{10pt}{$\scriptstyle{\rm (a)}$}
\includegraphics*[scale=.6, clip=false]{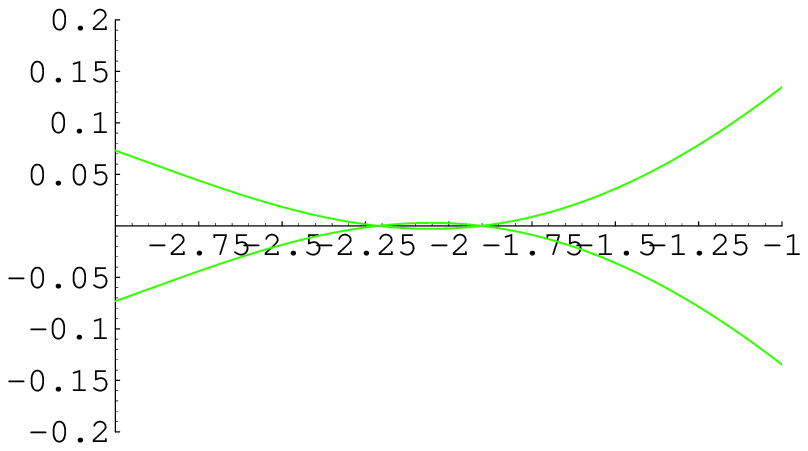}\quad
\raisebox{10pt}{$\scriptstyle{\rm (b)}$}
\includegraphics*[scale=.6, clip=false]{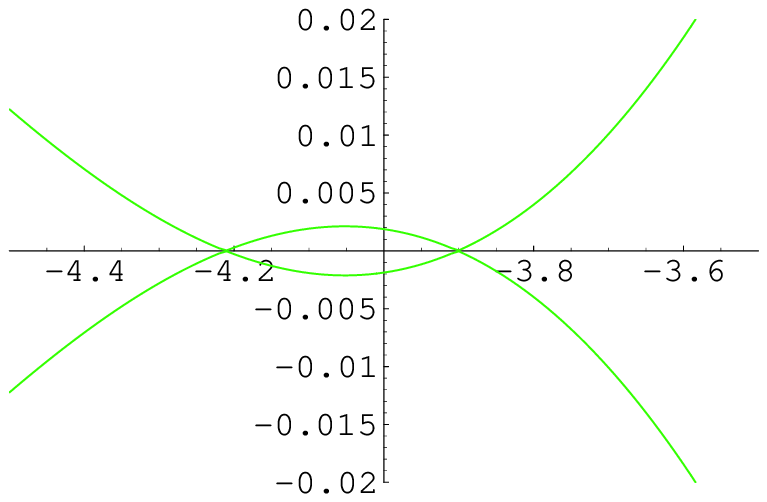}\\
\bigskip
\raisebox{10pt}{$\scriptstyle{\rm (c)}$}
\includegraphics*[scale=.6, clip=false]{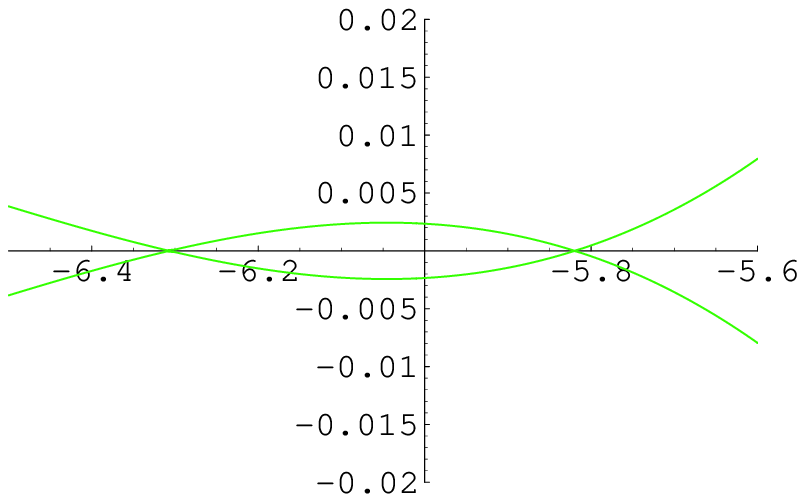}\quad
\end{center}
\caption[x] {\footnotesize The vanishing eigenvalue of the kinetic
operator for ghost number 1, at level 10 (or 9) in the scalar odd sector
(a), vector even (b) and vector odd (c) }
\label{f:g1double}
\end{figure}

For scalars in the odd sector and vectors in both odd and even sectors
there exists a first group of zeros on the negative $p^2$ axis which are
closest to $p^2=0$ and well separated from other zeros located at more
negatives values of $p^2$ (Graphs (b),(c),(d) of Figure
\ref{f:zerospdf}). These groups of ``almost degenerate'' zeros become
stable starting with level $L=4$ or $L=5$.  As the level increases these
zeros move on the $p^2$ axis but their number does not jump.  The almost
degenerate zeros of the scalar odd determinant are located around
$p^2=-2$; those of the vector even determinant around $p^2=-4$; and those
of vector odd determinant around $p^2=-6$.  In all these three cases, the
almost degenerate zeros form a reducible representation of the $SU(1,1)$
which is the sum of a scalar with $J=0$ , two doublets with $J=1/2$ and a
vector with $J=1$. The associated Fadeev-Popov index vanishes
\be
\sum_{J=0,1/2,1} (-1)^{2\,J} (2\,J+1)\, d_J =0
\ee
The observation which is important for our analysis is the following: the
two zeros with $J=1/2$ do correspond to a {\it single} eigenvalues of the
kinetic operator with a zero of order two. This seems to be unequivocal
looking at the graphs of the vanishing eigenvalue that is reported, for
level 10 or 9 in Figure \ref{f:g1double}.  The conclusion is that the two
doublets with $J=1/2$ should correspond in the exact theory to a single
zero with $d_J=2$.

The almost degenerate zeros of the determinants for the scalar odd, vector
even and vector odd sectors, are plotted, with the corresponding levels, in
Figures~\ref{f:scalarodd},\ref{f:vectoreven},\ref{f:vectorodd}. In the
same plots we also show the linear fits of the zeros locations  as function of the inverse of the level, $1/L$, for the different $SU(1,1)$ spins $J$.

The linearly extrapolated values of the zeros of the determinants for 
spins $J=0,1/2,1$ in the various Lorentz/twist-parity sectors are listed in Table
\ref{tabellazeri}.\footnote{The linear fits in
Figures~\ref{f:scalarodd},\ref{f:vectoreven},\ref{f:vectorodd} have been
performed by excluding the values of the zeros with lowest level, for
which one can expect the corrections to the linear dependence in $1/L$ are
largest. Including these zeros in the fits does not change the
extrapolated values in Table \ref{tabellazeri} significantly: it worsen
slightly the convergence between zeros with different ghost number.}

\begin{table}[ht]
\caption{Determinant zeros extrapolated at $L=\infty$}
\begin{center}
\begin{tabular}{|c|c|c|c|}
\hline
Sector &J=0 & J=1/2 &J=1\\ \hline
scalar odd & -1.99172&-2.03279;\, -1.97541&-2.04905\\ \hline
vector even &-3.98938 &-3.99494;\,  -3.99087 &-3.98803\\ \hline
vector odd & -5.97751&-5.96576;\, -6.00275&-5.78701\\ \hline
\end{tabular}
\end{center}
\label{tabellazeri}
\end{table}%

Extrapolated zeros with different $J$ agree with remarkable accuracy.  It
is very tempting to conjecture from these data that the {\it exact} values
for the degenerate zeros in the corresponding sectors are
\bea
&&  {m^2}_{\sss scalar, -} =2.0\nonumber\\
&&   {m^2}_{\sss vector, +}=4.0\nonumber\\
&& {m^2}_{\sss vector, -}=6.0 \nonumber
\eea

\sectiono{The action of $\Qtilde$ on the zeros of $\Ltilde$}
\label{sec:qaction}

We found, numerically, that the kernels $ \Wtilde{n}$ of
$\Lhat^{(n)}$ corresponding to multiplets of approximately degenerate
propagators poles, form a singlet, a doublet and a triplet of the $SU(1,1)$
symmetry, in all sectors listed in Table \ref{tabellazeri}. 
Let us denote by $v_{\pm2}$ and $v_{\pm1}$ the vectors that
generate, respectively, the kernels $ \Wtilde{\pm2}$ and $
\Wtilde{\pm1}$. Let $v_0^{s}$ and $v_0^{t}$ be the vectors in $
\Wtilde{0}$ that belong to, respectively, the singlet and the triplet of
$SU(1,1)$. One has
\bea 
&&J_+ \, v_0^{s}=0\quad v_0^{t}= J_+\,
v_{-2}\quad v_2= J_+ \,v_0^{t}\quad J_+\, v_2 =0\nonumber\\ 
&&v_1 = J_+\,v_{-1}\qquad J_{+}\, v_1=0 
\eea 
$J_+$ commutes both with $\Ltilde$ and with $\Qtilde$. It follows that it
commutes with $\Mhat$, $\Lhat$, $\Dhat$, $\Zhat$ and $ \Zcheck$.
Therefore $J_+$ maps not only $\Wtilde{n}$ into $\Wtilde{n+2}$ but also
$\Wcheck{n}$ into $\Wcheck{n+2}$.

The goal of this Section is to evaluate the dimensions of the tilde and
check relative cohomologies
\be
\n{n} \equiv \dim \htilde{n}\qquad  \ncheck{n} \equiv \dim \hcheck{n}
\ee  
As explained above, our numerical computations indicate that the
dimensions of the kernels $ \Wtilde{n}$ of
$\Lhat^{(n)}$ in the {\it exact} theory are
\be
\dim \Wtilde{0} = 2 \quad \dim \Wtilde{\pm 1} = 1
\quad \dim \Wtilde{\pm 2}= 
1\quad \dim \Wtilde{\pm n} = 0\quad {\rm for}\; n\geq 3
\label{experiment1}
\ee
Therefore, for the same values of $p^2$, the relative indices  are
\be
\mathrm{index} \,\tilde{\mathrm{h}}=\mathrm{index}\,
\check{\mathrm{h}}=\sum_n (-1)^n \dim\Wtilde{n} = 2
\label{experiment2}
\ee
while the Fadeev-Popov index vanishes and
\be
\H0(\Qtilde)= \H1(\Qtilde) = 0
\ee
in agreement with Sen's conjecture. When $\H0(\Qtilde)= \H1(\Qtilde) = 0$
the long sequence (\ref{nonpertbottsequence}) breaks into the short exact
sequence
\be
0\; {\mathop{\longrightarrow}^{\pi}}
\;\hcheck{-1}\;{\mathop{\longrightarrow}^{\Dcaltilde}}\;\htilde{1}\;
{\mathop{\longrightarrow}^{\imath}}\; 0
\label{bottshort}
\ee
and into the  two semi-infinite exact sequences
\bea
&&\!\!\!\!\!\!\!\!\cdots\;{\mathop{\longrightarrow}^{\pi}}\;\hcheck{-3}\;
{\mathop{\longrightarrow}^{\Dcaltilde}}\; \htilde{-1}\;
{\mathop{\longrightarrow}^{\imath }}\;\H{-1}(\Qtilde)\;
{\mathop{\longrightarrow}^{\pi}}
\;\hcheck{-2}\;{\mathop{\longrightarrow}^{\Dcaltilde}}\;\htilde{0}\;
{\mathop{\longrightarrow}^{\imath}}\;0\nonumber\\
&&\!\!\!0\;{\mathop{\longrightarrow}^{\pi}}\;\hcheck{0}\;
{\mathop{\longrightarrow}^{\Dcaltilde}}\; \htilde{2}\;
{\mathop{\longrightarrow}^{\imath }}\;\H{-1}(\Qtilde)\;
{\mathop{\longrightarrow}^{\pi}}
\;\hcheck{1}\;{\mathop{\longrightarrow}^{\Dcaltilde}}\;\htilde{3}\;
{\mathop{\longrightarrow}^{\imath}}\;\cdots\nonumber\\
\label{semibottsequence1}
\eea
From (\ref{bottshort}) one obtains
\be
\htilde{1}=\hcheck{-1}
\label{relation1}
\ee
and this should hold for {\it any} $p^2$ --- if Sen's conjecture is true.

Eq. (\ref{experiment1}) implies that the semi-infinite exact sequences
(\ref{semibottsequence1}) break up into finite sequences
\bea
&& 0\; 
{\mathop{\longrightarrow}^{\Dcaltilde}}\; \htilde{-1}\;
{\mathop{\longrightarrow}^{\imath }}\;\H{-1}(\Qtilde)\;
{\mathop{\longrightarrow}^{\pi}}
\;\hcheck{-2}\;{\mathop{\longrightarrow}^{\Dcaltilde}}\; \htilde{0}\;
{\mathop{\longrightarrow}^{\imath}}\;0\nonumber\\
&&0\;{\mathop{\longrightarrow}^{\pi}}\;\hcheck{0}\;
{\mathop{\longrightarrow}^{\Dcaltilde}}\; \htilde{2}\;
{\mathop{\longrightarrow}^{\imath }}\;\H{2}(\Qtilde)\;
{\mathop{\longrightarrow}^{\pi}}
\;\hcheck{1}\;{\mathop{\longrightarrow}^{\Dcaltilde}}\; 0\nonumber\\
&&0\;{\mathop{\longrightarrow}^{\Dcaltilde}}\;\htilde{-2}\;
{\mathop{\longrightarrow}^{\imath
}}\;\H{-2}(\Qtilde)\;{\mathop{\longrightarrow}^{\pi}}\; 0\nonumber\\
&&0\;{\mathop{\longrightarrow}^{\imath}}\;\H{3}(\Qtilde)\;
{\mathop{\longrightarrow}^{\pi}}\;\hcheck{2}
\;{\mathop{\longrightarrow}^{\Dcaltilde}}\;0 
\label{experiment3}
\eea
Hence, we obtain 
\be
\H{-2}(\Qtilde)=\H{3}(\Qtilde)=\htilde{-2}=\hcheck{2}
\label{experiment4}
\ee
from the last two sequences above, while the first two give
\be
\dim \H{-1}(\Qtilde) =
\n{-1}+
\ncheck{-2}- \n{0}=\dim \H{2}(\Qtilde) = - \ncheck{0}+ \n{2}+\ncheck{1}
\label{experiment5}
\ee
Eqs. (\ref{relation1}), (\ref{experiment4}), and (\ref{experiment5})
establish the following relations between the dimensions of the relative
tilde and check cohomologies:
\be
 \ncheck{-1}=\n{1}\qquad \ncheck{0}= 2+ \n{1}
-\n{-2}+\ncheck{1}-\ncheck{-2}\qquad \ncheck{2} = \n{-2}
 \label{tildecheck}
\ee
We now want to look for solutions of the equations
(\ref{experiment2}-\ref{experiment5}) with
\be
\n0, \ncheck0 =0,1,2, \qquad \n1, \ncheck1, \n2, \ncheck2 =0,1
\label{ranges}
\ee
Moreover, relations (\ref{nless}) and (\ref{ngreater}) require that 
\be
\ncheck{-2} \leq \n{-2}\quad {\rm and} \quad \ncheck2 \geq \n2
\label{inequalities}
\ee
\bigskip
The most general action of $\Mhat$ on the kernel of $\Lhat$ takes the form
\bea
&&\Mhat\, v_{-2} =\mu\, v_{-1}\qquad\qquad \Mhat\, v_{2} =0\nonumber\\ 
&&\Mhat\, v_{-1} =\alpha\,v_0^{s}+\beta\,v_0^{t}\qquad \Mhat\,v_{1} =
\beta\, v_{2}\nonumber\\
&&\Mhat\, v_0^{s} = \gamma\, v_1\qquad \Mhat\, v_0^{t}= \Mhat\, J_+ v_{-2}
=\mu\, v_{1}
\label{mhatfirst}
\eea
where $\mu$, $\alpha$, $\beta$ and $\gamma$ are numbers. (Without loss of
generality we can assume these numbers to be real, since their phases can
be reabsorbed into the normalizations of the vectors $v_{-n}$).
$\Mhat^2=0$ is equivalent to the relations
\bea
&&\mu\,\alpha=\mu\,\beta=0\nonumber\\
&&\alpha\,\gamma=\beta\,\gamma=0
\label{mhatsquare}
\eea
The possible solutions of (\ref{mhatsquare}) are

\bigskip

a) $\alpha=\beta=\gamma=\mu =0$

b) $\mu\not=0$, $\alpha=\beta=0$

c) $\mu=0$,  $\gamma=0$, $(\alpha,\beta) \not = (0,0)$

d) $\mu =\alpha=\beta =0$, $\gamma\not=0$

\bigskip

Only d), however, is compatible with the constraints that come from the
sequence (\ref{experiment3}) and the fact that $J_+$ commutes with
$\Qtilde$.  Let us show why.

Solution a)  leads to
\be
 \n{-2}= 1\quad \n{-1} =1\quad\n{0}=2\quad \n{1}=1\quad \n{2}=1
 \label{abczero}
 \ee
Since $\n{0}-\n{-1}=1$ it follows from (\ref{experiment5}) that 
\be
\ncheck{-2}=1 \qquad \dim \H{-1}(\Qtilde) =0
\label{n0n1uno}
\ee
If $\H{-1}(\Qtilde) =0$ the first two sequences in (\ref{experiment3})
split farther and give
\be
\ncheck{1}=\n{-1}=0
\label{n0n1unobis}
\ee
in conflict with Eq. (\ref{abczero}) above. 

Solution b) leads to the following values for the tilde cohomologies
\be
\n{-2}= 0 \quad \n{-1}=0 \quad \n{0}=1\quad \n{1} =0\quad \n{2}=1
\ee 
Again, $\n{0}-\n{-1}=1$ and therefore, as in (\ref{n0n1uno})
$\ncheck{-2}=1$.  But this is inconsistent with the inequalities
(\ref{inequalities}) which require $\ncheck{-2}\le \n{-2}=0$.

Two different sets of values for tilde cohomologies are a priori possible
in case of solution c):
\be
\n{-2}= 1 \quad \n{-1}=0 \quad \n{0}=1\quad \n{1} =n\quad \n{2}=n
\label{ctildesolution}
\ee 
with $n=1$ if $\beta=0$ and $n=0$ if $\beta\not=0$. In both cases,
$\n{0}-\n{-1}=1$, and therefore Eqs. (\ref{n0n1uno}) and
(\ref{n0n1unobis}) hold as well. Therefore the values of the check
cohomologies are
 \be
\ncheck{-2}= 1 \quad \ncheck{-1}=n \quad \ncheck{0}=n\quad \ncheck{1}
=0\quad \ncheck{2}=1
\label{cchecksolution}
\ee 
However, $n=1= \ncheck{-1}$ requires $\alpha=\beta=0$, otherwise $\Mhat$
would have no kernel on $\Wcheck{-1}\subset \Wtilde{-1}$.  This reduces
again to solution a), which we have already ruled out. Therefore $n=0$.
Since $\ncheck{-2}= 1$, $v_{-2} \in \Wcheck{-2}$.  Thus $v_0^t\in
\Wcheck{0}$ and $v_2\in \Wcheck{2}$.  $\ncheck{0}=0$ dictates that $v_0^t$
be trivial in check cohomology.  Therefore, $\alpha=0$ and
\be
\Mhat\, v_{-1} = \beta \, v_0^t
\ee
where $\beta\not=0$ and $v_{-1}\in \Wcheck{-1}$. As $J_+$ sends
$\Wcheck{-1}$ to $\Wcheck{1}$ we conclude that
\be
v_1 = J_+\, v_{-1} \in \Wcheck{1}
\ee
We reached a contradiction: $v_{2}\in \Wcheck{2}$, $v_{1}\in \Wcheck{1}$
and $\beta\,v_2 = \Mhat\, v_1$ means that $\ncheck{2}=0$, in conflict with
(\ref{cchecksolution}).

We are left therefore with solution d), for which 

\be
\n{-2}= 1 \quad \n{-1}=1 \quad \n{0}=1\quad \n{1} =0\quad \n{2}=1
\label{dtildesolution}
\ee 
The first sequence in (\ref{experiment3}) implies that  $\ncheck{-2}$ does not vanish --- since $\n{0}=1$. Thus $\ncheck{-2}=1$. Therefore
$v_{-2}\in \Wcheck{-2}$ and $v_0^t =J_+\, v_{-2} \in \Wcheck{0}$.  Since
$\Mhat\, v_0^t=0$ and $\Mhat\, v_{-1}=$, we conclude that $\ncheck{0}=1$.
This, together with the general relations (\ref{tildecheck}), determines
all of the check cohomologies:
\be
\ncheck{-2}= 1 \quad \ncheck{-1}=0 \quad \ncheck{0}=1\quad \ncheck{1}
=1\quad \ncheck{2}=1
\label{dchecksolution}
\ee 

To sum up, the action of the operators $\Mhat$ on the kernel $\Wtilde{n}$
and the subspaces $\Wcheck{n}$ which are compatible with our sequence are
\bea
&&\Mhat\, v_{\pm 2} =0\qquad\Mhat\, v_{\pm1} =0\qquad \Mhat\, v_0^{t}=0
\qquad \Mhat\, v_0^{s} = v_1\nonumber\\ 
&&\Wcheck{\pm 2}=\{v_{\pm 2}\}\quad \Wcheck{-1} =0\quad\Wcheck{1}=
\{v_{1}\}\quad \Wcheck{0}=\{ v_0^{t}\}
\label{maction}
\eea
Non-trivial representatives of $ \H{-1}(\Qtilde)$ and $ \H{-2}(\Qtilde)$
are $v_{-1}$ and $v_{-2}$, respectively. The dual non-empty cohomologies $
\H{2}(\Qtilde)$ and $ \H{3}(\Qtilde)$ have representatives
\be
\phi_2 = c_0\, v_1 + \phi_2^\prime\ee
and
\be
\phi_3 = c_0\, v_2 + \phi_3^\prime\ee
respectively, where $\phi_2^\prime$ and $\phi_3^\prime$ are given by
\be
\quad \Zhat\,v_1 = \Lhat\, \phi^\prime_2\qquad\quad \Zhat\,v_2 = \Lhat\,
\phi^\prime_3
\ee
Note that $\phi^\prime_2$ and $\phi^\prime_3$ are defined by these
equations up to elements in the kernels of $\Lhat$ on $\Wtilde{2}$ and
$\Wtilde{3}$. The latter is empty and the former is spanned by $v_2$: the
exactness of the sequence (\ref{experiment3}) ensures that $v_2$ is
trivial in the $\Qtilde$ cohomology.

\section*{Acknowledgments} 

I thank Sofia Mosci for collaborating on setting up the low-level
numerical computations. I thank Stefano Giusto and Martin Schnabl for
simulating conversations. I am indebted to M. Beccaria for useful
suggestions regarding computational algorithms.  I gratefully acknowledge the
hospitality of the Theory Group of CERN, where part of this work was
done. This work is supported in part by Ministero dell'Universit\`a e
della Ricerca Scientifica e Tecnologica.

\end{document}